\documentclass[prd,aps,twocolumn,preprintnumbers, showpacs, nofootinbib,superscriptaddress,notitlepage]{revtex4-1}
\usepackage{amsmath,graphicx,amssymb}
\usepackage{bm}
\usepackage{slashed}
\usepackage{xcolor}
\usepackage[colorlinks=true,linkcolor=blue,citecolor=blue,urlcolor=blue]{hyperref}

\allowdisplaybreaks

\newcommand{\nn}{\nonumber}

\begin{document}

\title{Mellin Moments of Proton Unpolarized GPDs at Nonzero Skewness \\ from Lattice QCD with a Neural-Network Parametrization}

\author{Min-Huan Chu}
\email{minhuan.chu@amu.edu.pl}
\affiliation{Faculty of Physics and Astronomy, Adam Mickiewicz University, ul.\ Uniwersytetu Pozna\'nskiego 2, 61-614 Pozna\'{n}, Poland}

\author{Manuel Cola\c{c}o}
\email{mancol@amu.edu.pl}
\affiliation{Faculty of Physics and Astronomy, Adam Mickiewicz University, ul.\ Uniwersytetu Pozna\'nskiego 2, 61-614 Pozna\'{n}, Poland}

\author{Krzysztof Cichy}
\affiliation{Faculty of Physics and Astronomy, Adam Mickiewicz University, ul.\ Uniwersytetu Pozna\'nskiego 2, 61-614 Pozna\'{n}, Poland}

\author{Martha Constantinou}
\affiliation{Department of Physics,  Temple University,  Philadelphia,  PA 19122 - 1801,  USA}

\author{Arif Sarkar}
\affiliation{Faculty of Physics and Astronomy, Adam Mickiewicz University, ul.\ Uniwersytetu Pozna\'nskiego 2, 61-614 Pozna\'{n}, Poland}

\begin{abstract}
We use lattice QCD to determine the generalized form factors (GFFs) entering the first four Mellin moments of the unpolarized isovector (\(u-d\)) proton GPDs at nonzero skewness. The framework combines the polynomiality of the moments in the skewness $\xi$, with the short-distance expansion formalism, allowing the lattice-extracted reduced Ioffe-time distributions (ITDs) to be expressed in terms of GFFs. In order to extract the GFFs by fitting these relations to the lattice data, we parameterize each GFF with an artificial neural network (ANN) as a continuous function of the invariant momentum transfer squared $t$, which accommodates the kinematic coverage of the data in a flexible way. With this fit, we determine the fourteen GFFs that enter the first four Mellin moments, including the skewness-dependent GFFs. From the extracted GFFs, we obtain the quark total angular momentum of the isovector combination, $J^{u-d}=0.296(11)(7)$ in the \(\overline{\mathrm{MS}}\) scheme at \(\mu=2~\mathrm{GeV}\), in agreement with previous lattice determinations based on nonlocal operators and with a smaller total uncertainty. We also determine the impact-parameter distributions of the first three Mellin moments in an unpolarized and in a transversely polarized proton. We finally identify and discuss the principal systematic uncertainties which enter this extraction, and the challenges specific to the nonzero-skewness regime.
\end{abstract}

\maketitle

\section{Introduction}
\label{sec:intro}
Understanding the internal structure of the nucleon has been one of the key aims of hadronic physics since deep inelastic scattering experiments revealed its partonic substructure almost 60 years ago. This structure is governed by quantum chromodynamics (QCD), the theory of the strong interaction within the Standard Model. QCD describes the interaction of color-charged quarks via gluons, which themselves carry color charge and therefore interact with each other. This self-interaction drives the complexity of the strong interaction and, consequently, the rich structure of all hadrons. As a first approximation, the nucleon can be viewed as a bound state of three valence quarks. In QCD, however, its full quantum state also contains dynamical quark--antiquark and gluon degrees of freedom. These constituents, collectively called partons, contribute to the nucleon's properties, including global ones such as its emergent mass and spin. Despite decades of theoretical and experimental effort, we still lack a complete quantitative understanding of the nucleon's structure.

The quantitative description of the nucleon's structure involves the determination of several partonic functions, the simplest of which are the one-dimensional parton distribution functions (PDFs) and form factors (FFs). PDFs describe the longitudinal momentum distribution of partons, expressed through their dependence on the momentum fraction $x$ carried by a parton. Different PDFs correspond to different polarizations of the partons and of the nucleon (unpolarized, helicity, and transversity PDFs) and to different twist, which determines the power of the hard scale with which they contribute to high-energy scattering. Experimentally, PDFs are extracted from inclusive high-energy processes, in particular deep inelastic scattering (DIS), in which the nucleon breaks up. Form factors provide a complementary picture: through their dependence on the momentum transfer squared $t$, they map out the spatial distribution of the nucleon's properties. The electromagnetic FFs describe the distribution of charge and magnetization and are measured in elastic scattering, where the nucleon remains intact. The gravitational form factors, in turn, parameterize the matrix elements of the energy-momentum tensor and encode the spatial distribution of the nucleon's mass and angular momentum, as well as its ``mechanical'' properties, such as the pressure and shear-force distributions~\cite{Polyakov:2002wz,Polyakov:2002yz,Polyakov:2018zvc}. Since the nucleon cannot be probed directly with gravitons, these FFs cannot be measured in elastic scattering. 

Generalized parton distributions (GPDs), introduced in the 1990s \cite{Mueller:1998fv,Ji:1996ek,Ji:1996nm,Radyushkin:1996nd}, provide a synthesis of PDFs and FFs. They depend on $x$, on $t$, and on the skewness $\xi$, which describes the longitudinal momentum transfer. The one-dimensional distributions are recovered from GPDs as limits or integrals: PDFs are obtained in the forward limit, $t,\,\xi\rightarrow0$, while the Mellin moments, namely the integrals of GPDs weighted with integer powers of $x$, yield form factors, the lowest moments giving the electromagnetic FFs and the second moments the gravitational ones. Importantly, some GPDs have no PDF counterpart, so they contain information not accessible in the forward limit. In particular, they allow for a three-dimensional imaging of the nucleon, depicting the distribution of partons in the plane transverse to its motion \cite{Burkardt:2000za,Ralston:2001xs,Diehl:2002he,Burkardt:2002hr}, and, through Ji's sum rule, they give access to the total angular momentum of quarks, which is indispensable for understanding the nucleon's spin structure~\cite{Ji:1996ek}. For more information on the physics of GPDs, see, e.g., the reviews~\cite{Ji:1998pc,Radyushkin:2000uy,Diehl:2003ny,Belitsky:2005qn,Boffi:2007yc,Guidal:2013rya,Lorce:2025aqp}.

GPDs are accessible in hard exclusive processes such as deeply virtual Compton scattering (DVCS) \cite{Mueller:1998fv,Ji:1996nm,Radyushkin:1996nd,Collins:1998be} and deeply virtual meson production (DVMP) \cite{Radyushkin:1996ru,Collins:1996fb,Mankiewicz:1997uy}.
Despite their theoretical appeal, experimentally extracting GPDs presents formidable challenges. While PDFs enter inclusive cross sections through a convolution with perturbatively calculable coefficient functions, GPDs enter the amplitudes of exclusive processes through complex-valued convolutions over $x$. In practice, experimental data constrain GPDs mainly on or close to the line $x=\pm\xi$, which makes a full three-dimensional reconstruction from experimental data alone an ill-posed inverse problem that requires model-dependent assumptions. Processes such as double DVCS (DDVCS) \cite{Belitsky:2002tf,Guidal:2002kt,Deja:2023ahc} can overcome the restriction to $x=\pm\xi$ by varying the virtualities of the incoming and outgoing photons independently, but their cross sections are strongly suppressed, and they may become accessible only in the future, e.g.\ at the Electron-Ion Collider (EIC) and at Jefferson Lab.

These difficulties make lattice QCD (LQCD) an important complementary tool for the study of GPDs. By simulating the strong interaction on a discretized Euclidean spacetime, LQCD allows one to compute from first principles the non-perturbative matrix elements from which GPDs can be extracted. Initially, lattice studies focused on Mellin moments of GPDs obtained from matrix elements of local operators via the operator product expansion (OPE), see, e.g., Refs.~\cite{Shintani:2018ozy,Bali:2018zgl,Alexandrou:2018sjm,Alexandrou:2019olr,Alexandrou:2019ali,Alexandrou:2020okk,Alexandrou:2021wzv,Alexandrou:2023qbg,Hackett:2023rif,Alexandrou:2025vto,Alexandrou:2026tjs}. While these studies provided important insight into the structure of hadrons, they cannot give access to the $x$ dependence of GPDs, since in practice only the lowest few moments can be computed: operators with a larger number of covariant derivatives mix, due to the reduced rotational symmetry of the lattice, with lower-dimensional operators, and their signal deteriorates rapidly.

More recently, a new route to partonic distributions has been proposed, based on non-local matrix elements of boosted hadron states. Since LQCD cannot probe correlations on the light front, the access to GPDs via this route is indirect: matrix elements of operators with space-like separations are used to construct observables that can be factorized onto the desired partonic functions using perturbation theory. The breakthrough was the proposal by Ji \cite{Ji:2013dva,Ji:2014gla} of quasi-distributions and their matching to light-cone distributions within Large-Momentum Effective Theory (LaMET), with factorization performed in momentum space. Subsequently, Radyushkin proposed an alternative approach based on the same non-local matrix elements, the pseudo-distributions, with factorization performed in coordinate space (short-distance factorization, SDF) \cite{Radyushkin:2017cyf}. Since then, a number of lattice studies of GPDs have been carried out, see, e.g., Refs.~\cite{Chen:2019lcm,Alexandrou:2020zbe,Alexandrou:2021bbo,Lin:2020rxa,Lin:2021brq,Lin:2023gxz,Holligan:2023jqh,Bhattacharya:2023nmv,Bhattacharya:2022aob,Cichy:2023dgk,Bhattacharya:2023ays,Bhattacharya:2023jsc,Bhattacharya:2024wtg,Ding:2024saz,Bhattacharya:2024qpp,Cichy:2024afd,Chu:2025kew,HadStruc:2024rix,Burkert:2025gzu,Chu:2025jsi,Gao:2025inf,Dutrieux:2026grg}. Further alternative approaches exist \cite{Liu:1993cv,Detmold:2005gg,Braun:2007wv,Chambers:2017dov,Ma:2017pxb,Francis:2025rya}, with explicit applications to GPDs \cite{CSSMQCDSFUKQCD:2021lkf,Liang:2023uai,Hannaford-Gunn:2024aix}. For more information on the progress of LQCD calculations of partonic distributions, we refer to several reviews, e.g., Refs.~\cite{Lin:2017snn,Cichy:2018mum,Radyushkin:2019mye,Ji:2020ect,Constantinou:2020pek,Cichy:2021lih,Cichy:2021ewm,Constantinou:2022yye,Burkert:2022hjz,Lin:2025hka,Boer:2025ixc,Constantinou:2026wii,Cichy:2026xrh}.

In this paper, we use SDF to determine the generalized form factors (GFFs) that enter the Mellin moments of the unpolarized GPDs at zero and nonzero skewness. Lattice matrix elements, renormalized in a ratio scheme, can be related to the GFFs via Wilson coefficients computed in perturbation theory. This approach was first applied to moments of PDFs in Ref.~\cite{Karpie:2018zaz} and later extended to moments of zero-skewness unpolarized~\cite{Bhattacharya:2023ays} and helicity~\cite{Bhattacharya:2024wtg} GPDs. Mellin moments, including their skewness dependence, were extracted within SDF in Refs.~\cite{HadStruc:2024rix,Gao:2025inf,Dutrieux:2026grg}. All these works employ the framework of asymmetric frames of reference \cite{Bhattacharya:2022aob,Bhattacharya:2023jsc}, which allows for a computationally efficient probing of a wide range of momentum transfers and skewness values. Within the same framework, an alternative, Lorentz-invariant definition of quasi-GPDs was introduced and shown to improve the convergence of the quasi-$E$ GPD to its light-cone counterpart~\cite{Bhattacharya:2022aob}.

The present work extends these studies in two directions. First, we add a new data set at a larger boost momentum, which, together with the data in Ref.~\cite{Chu:2025kew}, provides sufficient kinematic coverage. Second, and most importantly, we parametrize the $t$ dependence of each GFF with an artificial neural network (ANN), without assuming an analytical form. This is particularly important in asymmetric-frame kinematics, where the same spatial momentum-transfer vector corresponds to different $t$ values at different proton momentum boosts and skewness values. With this approach, we determine fourteen GFFs in total that enter the first four Mellin moments, including the skewness-dependent ones. We also evaluate the total angular momentum of quarks from Ji's sum rule, finding agreement with previous lattice determinations and a smaller total uncertainty, as well as the impact-parameter distributions of the first three Mellin moments. Finally, we discuss in detail the systematic uncertainties of the extraction. The present calculation is limited by the available kinematic coverage and by several sources of systematic uncertainty. We therefore regard this study primarily as a proof of concept for a future precision determination.

The paper is organized as follows. In Section~\ref{sec:theory}, we outline the theoretical framework. Section~\ref{sec:ann} describes the neural-network parametrization of the $t$ dependence of the GFFs. Our main numerical results are presented in Section~\ref{sec:results}, and the systematic studies in Section~\ref{sec:systematics}. We conclude and discuss future prospects in Section~\ref{sec:conclusions}. Details of the RGR matching and additional systematic tests are collected in the Appendices.

\section{Theoretical Framework}
\label{sec:theory}
In this section, we first introduce the unpolarized GPDs and their Mellin moments, whose skewness dependence is constrained by polynomiality relations that are expressed in terms of GFFs. We then
present the short-distance expansion that connects the lattice quantities to the moments of GPDs. Combining these relations gives the expressions used to fit the GFFs to the lattice data.

\subsection{The definition of Mellin moments for GPDs}

The leading-twist unpolarized quark GPDs of the proton are defined from the Fourier transform of the matrix element of a quark bilinear whose fields are separated along the light cone \cite{Mueller:1998fv,Ji:1996ek,Radyushkin:1996nd,Ji:1998pc},
\begin{align}
\bar{F}^{q}&(x,\xi,t)
= \int \frac{{\rm d}z^{-}}{4\pi}\, e^{\,i x P^{+} z^{-}}
\nonumber\\
&\times
\langle P_f|\,\bar{\psi}^{q}\!\left(-\tfrac{z}{2}\right)\gamma^{+}
\hat{W}\!\left(-\tfrac{z}{2},\tfrac{z}{2}\right)
\psi^{q}\!\left(\tfrac{z}{2}\right)|P_i\rangle\Big|_{z^+=0,z_{\perp}=0_{\perp}}
\nonumber\\
&= \frac{1}{2P^{+}}\,\bar{u}(P_f)
\left[\gamma^{+} H^{q}(x,\xi,t)\right.
\nonumber\\
&\qquad\qquad\qquad\qquad\;\left.
+ \frac{i\sigma^{+\nu}\Delta_{\nu}}{2m}\, E^{q}(x,\xi,t)\right]
u(P_i),
\label{eq:gpd-def}
\end{align}
where \(q\) denotes the quark flavor, $\hat{W}$ is the Wilson line in the fundamental representation along the light-like direction, $m$ is the proton mass, and $P_{i}$, $P_{f}$ are the momenta of the initial and final states, whose spin indices are suppressed throughout. The kinematic variables entering Eq.~\eqref{eq:gpd-def} are defined as
\begin{align}
P^{\mu} \equiv \frac{1}{2}\left(P^{\mu}_i+P^{\mu}_f\right),\;
\Delta^{\mu} \equiv P^{\mu}_f-P^{\mu}_i,\;
t \equiv \Delta^{2},\;
\xi \equiv -\frac{\Delta^{+}}{2P^{+}} .
\label{eq:kinematics}
\end{align}
In the forward limit $\xi=0$, $t=0$, the GPD $H^{q}$ reduces to the unpolarized parton distribution function (PDF) $H^{q}(x,0,0)=q(x)$.

There are two main approaches to the study of GPDs in lattice QCD. The first is to compute the $x$ dependence of the GPDs \cite{Chu:2025kew}. The second one utilized in this work is to determine their Mellin moments, which are defined as
\begin{align}
\bar{F}^q_{n+1}(\xi,t) \equiv \int_{-1}^{1}{\rm d}x\;x^{n}\bar{F}^q(x,\xi,t),\; \bar{F}^q = H^{q},\,E^{q},
\label{eq:mellin-def}
\end{align}
where $\bar{F}_{n+1}$ represents the Mellin moment of $H^{q}$ or $E^{q}$. Lorentz covariance, together with time-reversal invariance, constrains these moments to be polynomials in even powers of the skewness
\cite{Ji:1998pc}. Explicitly,
\begin{align}
H_{n+1}(\xi,t) &= \sum_{\substack{i=0 \\ i\ {\rm even}}}^{n}
\xi^iA_{n+1,i}(t)+{\rm mod}(n,2)\,\xi^{\,n+1}\,C_{n+1}(t),
\label{eq:polyH}\\
E_{n+1}(\xi,t) &= \sum_{\substack{i=0 \\ i\ {\rm even}}}^{n}
\xi^{i}\,B_{n+1,i}(t)-{\rm mod}(n,2)\,\xi^{\,n+1}\,C_{n+1}(t),
\label{eq:polyE}
\end{align}
where the flavor index $q$ is omitted for simplicity and the coefficients $A_{n+1,i}(t)$, $B_{n+1,i}(t)$ and $C_{n+1}(t)$ are the GFFs. Note that the moments depend on both $\xi$ and $t$, whereas the GFFs are functions of $t$ only. We adopt the convention in which the expansion is performed in powers of $\xi$, so that the GFFs defined here differ by factors of $2^{i}$ from those of works expanding in powers of $2\xi$.

In this work we study the Mellin moments up to the fourth order ($n=3$). Writing out the first four orders, the accessible moments read
\begin{align}
n=0:\quad
&H_{1}=A_{1,0}(t),\nonumber\\
&E_{1}=B_{1,0}(t),
\label{eq:mom1}\\
n=1:\quad
&H_{2}=A_{2,0}(t)+\xi^{2}C_{2}(t),\nonumber\\
&E_{2}=B_{2,0}(t)-\xi^{2}C_{2}(t),
\label{eq:mom2}\\
n=2:\quad
&H_{3}=A_{3,0}(t)+\xi^{2}A_{3,2}(t),\nonumber\\
&E_{3}=B_{3,0}(t)+\xi^{2}B_{3,2}(t),
\label{eq:mom3}\\
n=3:\quad
&H_{4}=A_{4,0}(t)+\xi^{2}A_{4,2}(t)+\xi^{4}C_{4}(t),\nonumber\\
&E_{4}=B_{4,0}(t)+\xi^{2}B_{4,2}(t)-\xi^{4}C_{4}(t).
\label{eq:mom4}
\end{align}
While $H$ reduces to the PDF in the forward limit, the forward limit of the $E$-type GPD is not accessible on the lattice, because $E$ enters Eq.~\eqref{eq:gpd-def} multiplied by $\Delta_{\nu}$ and therefore does not contribute at $\Delta=0$. At $t=\xi=0$ the mixing terms, which carry a factor $\xi^{2}$, vanish as well. Thus, in the forward limit, the only surviving GFFs in Eqs.~\eqref{eq:mom1}--\eqref{eq:mom4} are \(A_{n+1,0}(0)\). 

The GFFs entering the lowest Mellin moments have a direct physical
interpretation. The $n=0$ GFFs correspond to the Dirac and Pauli
elastic form factors,
$A_{1,0}^q(t)=F_1^q(t)$ and $B_{1,0}^q(t)=F_2^q(t)$.
The $n=1$ GFFs, $A_{2,0}^q(t)$, $B_{2,0}^q(t)$, and $C_2^q(t)$,
are gravitational form factors, as they parameterize the nucleon
matrix element of the quark energy-momentum tensor
\cite{Polyakov:2002wz,Polyakov:2002yz,Polyakov:2018zvc}.
The GFFs $A_{2,0}^q$ and $B_{2,0}^q$ enter Ji's sum rule for the
total quark angular momentum \cite{Ji:1996ek},
\begin{align}
J^q
&=
\frac{1}{2}\int_{-1}^{1}{\rm d}x\;x
\left[
H^{q}(x,\xi,t=0)+E^{q}(x,\xi,t=0)
\right]
\nonumber\\
&=
\frac{1}{2}\Big[
A_{2,0}^q(0)+B_{2,0}^q(0)
\Big].
\label{eq:jisumrule}
\end{align}
The GFF $C_2^q(t)$ is related to the quark $D$-term form factor and is associated
with the ``mechanical'' properties of the nucleon.

This work aims to determine the full
polynomial structure of the first few Mellin moments, with special attention to the physically significant GFFs.
Computing the GFFs that only contribute at $n>1$ provides access to
the GFFs $A_{3,0}$, $B_{3,0}$, $A_{3,2}$, $B_{3,2}$, $A_{4,0}$, $B_{4,0}$, $A_{4,2}$, $B_{4,2}$, and $C_4$, therefore providing additional constraints on the lower GFFs with a
phenomenological interpretation. 

\subsection{Short distance approach}
\label{sec:sde}

The operator in Eq.~\eqref{eq:gpd-def} involves quark fields separated along a light-like direction and is therefore not accessible directly on a Euclidean lattice. It can be replaced by an equal-time operator with a space-like separation $z^{\mu}=(0,0,0,z)$, whose matrix elements can be computed on the lattice,
\begin{align}
M^{\mu}&(z,P,\Delta)\nonumber\\
&= \langle P_f|\,\bar{\psi}^{q}\!\left(-\tfrac{z}{2}\right)\gamma^{\mu}\hat{W}\!\left(-\tfrac{z}{2},\tfrac{z}{2}\right)\psi^{q}\!\left(\tfrac{z}{2}\right)|P_i\rangle .
\label{eq:bare-me}
\end{align}
We work throughout with the flavor-nonsinglet combination $u-d$, which receives no contributions from disconnected diagrams. The flavor index is omitted in the following discussions.

The space-like matrix element can be decomposed into eight Lorentz-invariant amplitudes $A_{k}$~\cite{Bhattacharya:2022aob,Chu:2025kew}. Writing
\begin{align}
M^{\mu}(z,P,\Delta) = \bar{u}(P_f)\,\tilde{M}^{\mu}(z,P,\Delta)\,u(P_i)\,,
\label{eq:ME-spinors}
\end{align}
we have
\begin{align}
\label{eq:MEparametrization}
\tilde{M}^{\mu}(z,P,\Delta)
&=
\dfrac{P^{\mu}}{m}A_1
+ m z^{\mu} A_2
+ \dfrac{\Delta^{\mu}}{m}A_3
+ i m \sigma^{\mu z} A_4
\nonumber\\
&\quad
+ \dfrac{i\sigma^{\mu\Delta}}{m}A_5
+ \dfrac{P^{\mu}i\sigma^{z\Delta}}{m}A_6
+ m z^{\mu}i\sigma^{z\Delta}A_7\nonumber\\
&\quad + \dfrac{\Delta^{\mu}i\sigma^{z\Delta}}{m}A_8\,,
\end{align}
with summation over repeated indices and the following compact notation:
$\sigma^{\mu \nu} \equiv \tfrac{i}{2} (\gamma^\mu \gamma^\nu - \gamma^\nu \gamma^\mu)$,
$\sigma^{\mu z} \equiv \sigma^{\mu \rho} z_\rho$,
$\sigma^{\mu \Delta} \equiv \sigma^{\mu \rho} \Delta_\rho$, $\sigma^{z \Delta} \equiv \sigma^{\rho \tau} z_\rho \Delta_\tau$.
The amplitudes are obtained from the projected matrix elements
\begin{align}
\Pi_{\mu}(\Gamma_{\kappa}) = K\,{\rm Tr}\left[\Gamma_{\kappa}\frac{-i\slashed{P}_{f}+m}{2m}\,\tilde{M}^{\mu}(z,P,\Delta)\,\frac{-i\slashed{P}_{i}+m}{2m}\right],
\label{eq:projected-me}
\end{align}
where $K=2m^{2}/\sqrt{E_{f}E_{i}(E_{f}+m)(E_{i}+m)}$, and the $\Gamma_{\kappa}$ are the parity projectors
\begin{align}
\Gamma_{0}=\frac{1}{4}\left(1+\gamma_{0}\right),\;\Gamma_{k}=\frac{1}{4}\left(1+\gamma_{0}\right)i\gamma_{5}\gamma_{k},
\label{eq:projectors}
\end{align}
the first one being unpolarized and the remaining three polarized along the direction $k$ ($k=1,2,3$). Combining the four insertions $\gamma_{\mu}$ with the four projectors in Eq.~\eqref{eq:projected-me} yields a linear system of equations for the eight amplitudes, which can be solved for them. More details on the extraction of the amplitudes are given in Ref.~\cite{Chu:2025kew}.

There are two Lorentz-invariant combinations of the amplitudes that converge to the GPDs $H^{q}$ and $E^{q}$~\cite{Bhattacharya:2022aob,Dutrieux:2024rem},
\begin{align}
\mathcal{F}^{H} &= A_{1}-2\xi\,A_{3},
\label{eq:ITD-H}\\
\mathcal{F}^{E} &= -A_{1}+2\xi\,A_{3}+2A_5+2P_3zA_6-4\xi P_3zA_8,
\label{eq:ITD-E}
\end{align}
where the skewness is defined as $\xi=(P^z_i-P^z_f)/(P^z_i+P^z_f)$. This space-like skewness equals its light-cone counterpart up to the difference from power corrections $\mathcal{O}(m^{2}/P_3^{2})$~\cite{Liu:2019urm,Bhattacharya:2019cme}. Eqs.~\eqref{eq:ITD-H} and \eqref{eq:ITD-E} combine bare coordinate-space matrix elements, which, when written in terms of the Ioffe time $\nu\equiv P_3 z$, are called Ioffe-time distributions (ITDs). These bare objects carry a logarithmic divergence and a power divergence generated by the self-energy of the Wilson line. Following Ref.~\cite{Radyushkin:2017cyf}, we define the ratio
\begin{align}
F(\nu,\xi,t,z^{2})
=
\frac{\mathcal{F}(\nu,\xi,t,z^{2})}{f(0,z)},
\label{eq:doubleratio}
\end{align}
where $\mathcal{F}=\mathcal{F}^{H},\mathcal{F}^{E}$ and $F=F^{H},F^{E}$, while $f(P_3,z)$ denotes the corresponding forward ($\Delta=0$) matrix element of the same $u-d$ operator. The factor $f(0,z)$ carries the same $z$-dependent multiplicative renormalization, including that associated with the Wilson line, as the nonforward matrix element, and therefore cancels these ultraviolet divergences in the ratio. The resulting reduced matrix element is free of multiplicative operator-renormalization factors. The factorization-scale dependence enters later through perturbative short-distance matching to the $\overline{\mathrm{MS}}$ scheme. The resulting functions \(F(\nu,\xi,t,z^2)\) are the reduced ITDs. In the following, we suppress the superscript distinguishing \(H\)- and \(E\)-type quantities unless explicitly needed.

Taylor expanding the exponential of the Fourier transform,
\begin{align}
F(\nu,\xi,t,z^{2}) = \int_{-1}^{1}{\rm d}x\; e^{-i\nu x}\,F(x,\xi,t,z^{2}),
\label{eq:FT}
\end{align}
and exchanging the sum with the integral over the bounded domain $x\in[-1,1]$ gives
\begin{align}
F(\nu,\xi,t,z^{2}) = \sum_{n=0}^{\infty}\frac{(-i\nu)^{n}}{n!}\,F_{n+1}(\xi,t,z^{2}),
\label{eq:nu-expansion}
\end{align}
with $F_{n+1}(\xi,t,z^{2})=\int_{-1}^{1}{\rm d}x\,x^{n}F(x,\xi,t,z^{2})$. Since $|x|\le1$ implies $|x|^{n}\le1$, the moments obey
\begin{align}
\left|F_{n+1}(\xi,t,z^{2})\right| \le \int_{-1}^{1}{\rm d}x\left|F(x,\xi,t,z^{2})\right| ,
\label{eq:bound}
\end{align}
that is, they are bounded at every $n$ by a quantity independent of $n$. The size of the neglected terms is therefore controlled by the factor $\nu^{n}/n!$ alone, and the validity of the truncation is a question of whether $|\nu|$ is small enough. We truncate the series at $n_{\rm max}=3$ and examine the effect of this choice in Sec.~\ref{sec:systematics}.

The operator product expansion relates the moments of the space-like correlator to the light-cone ones in a short-distance factorization relation \cite{Radyushkin:2019mye,Dutrieux:2024rem,HadStruc:2024rix},
\begin{align}
F_{n+1}(\xi,t,z^{2}) = \sum_{\substack{k=0 \\ k\ {\rm even}}}^{n}\xi^{k}\,c_{n+1,k}(z^{2}\mu^{2})\,\bar{F}_{n+1-k}(\xi,t),
\label{eq:matching-moments}
\end{align}
up to power corrections of $\mathcal{O}(z^{2}\Lambda_{\rm QCD}^{2})$, $\mathcal{O}(z^{2}m^{2})$ and $\mathcal{O}(z^{2}t)$~\cite{Braun:2026rgj}. Equation~\eqref{eq:matching-moments} relates space-like quantities on the left to the Mellin moments of GPDs on the right, where the Wilson coefficients $c_{n+1,k}$ encode the short-distance difference between these two functions and can be computed by perturbation theory. Only even powers of the skewness appear, as a consequence of Lorentz covariance and time-reversal invariance, and both the coefficients and the moments are real. The relation is thus diagonal at $\xi=0$, whereas at nonzero skewness the moments start to mix at order $\xi^{2}$, connecting, up to $n=n_{\rm max}=3$, $F_{3}$ with $\bar{F}_{1}$ and $F_{4}$ with $\bar{F}_{2}$,
\begin{align}
F_{1} &= \bar{F}_{1},\nonumber\\
F_{2} &= c_{2,0}\bar{F}_{2},\nonumber\\
F_{3} &= c_{3,0}\bar{F}_{3}+\xi^{2}c_{3,2}\bar{F}_{1},\nonumber\\
F_{4} &= c_{4,0}\bar{F}_{4}+\xi^{2}c_{4,2}\bar{F}_{2}.
\label{eq:match_mom}
\end{align}
\begin{table}[t]
\caption{Wilson coefficients $c_{n+1,k}(z^{2}\mu^{2})$ of Eq.~\eqref{eq:matching-moments} at order $\mathcal{O}(\alpha_{s})$ and in the $\overline{\rm MS}$ scheme, with $a_{s}=\alpha_{s}(\mu^{2})C_{F}/2\pi$ and $L=\ln\left(e^{2\gamma_{E}+1}z^{2}\mu^{2}/4\right)$, taken from Ref.~\cite{HadStruc:2024rix}.}
\label{tab:wilson}
\begin{ruledtabular}
\begin{tabular}{ccc}
$n$ & $c_{n+1,0}$ & $c_{n+1,2}$ \\
\colrule
$0$ & $1$ & \\
$1$ & $1-a_{s}\left(-\frac{4}{3}L+\frac{14}{3}\right)$ & \\
$2$ & $1-a_{s}\left(-\frac{25}{12}L+\frac{47}{6}\right)$ & $-a_{s}\left(\frac{5}{12}L-\frac{7}{6}\right)$ \\
$3$ & $1-a_{s}\left(-\frac{157}{60}L+\frac{931}{90}\right)$ & $-a_{s}\left(\frac{11}{20}L-\frac{53}{30}\right)$
\end{tabular}
\end{ruledtabular}
\end{table}

Truncating Eq.~\eqref{eq:nu-expansion} at $n_{\rm max}=3$ and inserting Eq.~\eqref{eq:match_mom} into it, we obtain
\begin{align}
{\rm Re}\,F &= \left(1-\frac{\nu^{2}}{2}\xi^{2}c_{3,2}\right)\bar{F}_{1}-\frac{\nu^{2}}{2}c_{3,0}\,\bar{F}_{3},
\label{eq:ReF}\\
{\rm Im}\,F &= -\nu\left(c_{2,0}-\frac{\nu^{2}}{6}\xi^{2}c_{4,2}\right)\bar{F}_{2}+\frac{\nu^{3}}{6}c_{4,0}\,\bar{F}_{4}.
\label{eq:ImF}
\end{align}
These relations, combined with the polynomiality Eqs.~\eqref{eq:mom1}--\eqref{eq:mom4}, yield the equations that can be used to fit the GFFs to the reduced ITD data, as discussed in Sec.~\ref{sec:ann-param}.

The coefficients $c_{n+1,k}$ from Table~\ref{tab:wilson} are evaluated at NLO in the $\overline{\rm MS}$ scheme, and in a first prescription, which we refer to as the fixed energy scale (FS) analysis, at the fixed scale $\mu=2\,{\rm GeV} $. For this work,
$\alpha_s(2\,{\rm GeV})\approx0.394$ \cite{FlavourLatticeAveragingGroupFLAG:2024oxs}. Using a Taylor expansion in $\alpha_s$, the diagonal coefficients can be written in the inverted form
\begin{align}
c_{n+1,0}&=\frac{1}{1+a_{s}(\gamma_{n+1}L+b_{n+1})}\nonumber\\
&=1-a_{s}\left(\gamma_{n+1}L+b_{n+1}\right)+\mathcal{O}(\alpha_{s}^{2}),
\label{eq:dn-fs}
\end{align}
where the coefficients $\gamma_{n+1}$ and $b_{n+1}$ entering $c_{n+1,0}$ are given in Table~\ref{tab:wilson}. This choice is equivalent to the tabulated expression to the order to which the coefficients are known.
The difference between the two forms starts at \(\mathcal{O}(\alpha_s^2)\) and therefore provides a measure of the sensitivity to neglected higher-order terms. We therefore test both prescriptions for $c_{n+1,0}$ and use their difference as an estimate of the perturbative truncation uncertainty (see Sec.~\ref{sec:sys-alpha}).

The coefficients in Table~\ref{tab:wilson} contain a logarithm \(L\) that can become large in magnitude, depending on the value of \(\mu^2 z^2\). These large logarithms can spoil the convergence of the perturbative series. In the renormalization group resummation (RGR) analysis, these logarithms are resummed at leading-logarithmic accuracy by using the combinations of moments that diagonalize the leading-order evolution of the GPDs. We collect the moments in the vectors $\bm{F}=(F_1,F_2,F_3,F_4)^{T}$ and $\bar{\bm{F}}=(\bar{F}_1,\bar{F}_2,\bar{F}_3,\bar{F}_4)^{T}$, and following Ref.~\cite{HadStruc:2024rix} the RGR relation takes the form
\begin{align}
\bar{\bm{F}} = G^{-1}(\xi)\,A(\mu^{2},z^{2})\,G(\xi)\left[I-B(\xi,z^{2})\right]\bm{F},
\label{eq:rgr}
\end{align}
where $B$ contains the one-loop coefficients of Table~\ref{tab:wilson} evaluated at the initial matching scale $\mu_{0}$, $G$ diagonalizes the leading-order evolution, and $A$ carries the resummed evolution from $\mu_{0}$ to $\mu$. The three matrices are collected in Appendix~\ref{app:matching_RGR}. We take the initial matching scale to be $\mu_{0}=2\kappa e^{-\gamma_{E}}/z$, with $\kappa$ a constant not fixed by the formalism, and write
\begin{align}
a_{s}=\frac{\alpha_{s}(\mu_{0}^{2})C_{F}}{2\pi},\;l=1+2\ln\kappa,\;r=\left[\frac{\alpha_{s}(\mu_{0}^{2})}{\alpha_{s}(\mu^{2})}\right]^{\frac{C_{F}}{2\pi\beta_{0}}},
\label{eq:rgr-shorthands}
\end{align}
with $\beta_{0}=(33-2n_{f})/(12\pi)$ and $\alpha_{s}$ computed at one loop. Inverting Eq.~\eqref{eq:rgr} gives the coefficients that replace those of the fixed-scale extraction in Eqs.~\eqref{eq:ReF} and \eqref{eq:ImF},
\begin{align}
c_{2,0} &= \frac{r^{4/3}}{d_{2}},\; c_{3,0} = \frac{r^{25/12}}{d_{3}},\; c_{4,0} = \frac{r^{157/60}}{d_{4}},
\label{eq:crgr-diag}\\
c_{3,2} &= \frac{1}{d_{3}}\left[\frac{1-r^{25/12}}{5}-a_{s}\left(\frac{5}{12}l-\frac{7}{6}\right)\right],
\label{eq:crgr-32}\\
c_{4,2} &= \frac{1}{d_{4}}\left[\frac{3\left(r^{4/3}-r^{157/60}\right)}{7}-\frac{r^{4/3}}{d_{2}}a_{s}\left(\frac{11}{20}l-\frac{53}{30}\right)\right],
\label{eq:crgr-42}
\end{align}
where 
\begin{align}
d_{n+1} = \left[1+a_{s}\left(\gamma_{n+1}\,l+b_{n+1}\right)\right].
\label{eq:dn}
\end{align}

To conclude this subsection, we note that many of the equations derived here are valid only under certain circumstances. Four conditions restrict the kinematics that can be used with Eqs.~\eqref{eq:ReF} and \eqref{eq:ImF}. First, the truncation of the expansion in Ioffe time requires $|\nu|$ to be small enough for the terms with $n\ge 4$ to be negligible at the present statistical precision. Second, the relation in Eq.~\eqref{eq:matching-moments} holds up to power corrections of order $z^{2}\Lambda_{\rm QCD}^{2}$, $z^{2}m^{2}$ and $z^{2}t$, which limits the size of both the separations $z$ and the momentum transfers $t$ that can be included. Third, the validity of the coefficients $c_{n+1,k}$ requires the perturbative expansion to converge, which again favors short separations. Fourth, at separations of order the lattice spacing, discretization effects affect the matrix elements that are not accounted for. We examine the systematic uncertainties associated with all these conditions in Sec.~\ref{sec:systematics}.

\section{Neural Network Parametrization of the $t$-dependent Generalized Form Factors}
\label{sec:ann}

 Eqs.~\eqref{eq:ReF} and \eqref{eq:ImF} show the \(\nu\)- and \(\xi\)-dependence of the reduced ITDs in terms of the GPD moments \(\bar F_{n+1}(\xi,t)\), while the corresponding Wilson coefficients are estimated from perturbation theory as outlined in Table~\ref{tab:wilson}. The quantities to be determined as functions of the Lorentz-invariant momentum transfer squared, $t$, are the GFFs for both $H$- and $E$-type GPDs. These consist of $A_{1,0}$, $A_{2,0}$, $A_{3,0}$, $A_{3,2}$, $A_{4,0}$, $A_{4,2}$ associated with $H$-type, and $B_{1,0}$, $B_{2,0}$, $B_{3,0}$, $B_{3,2}$, $B_{4,0}$, $B_{4,2}$ associated with $E$-type, along with $C_2$ and $C_4$, which are common to both types. 
 
 The kinematic variables $t$, $\xi$, and $\nu$ are correlated, as at least two of them always depend on the same set of variables. To be precise, changing either $P_{i,3}$ or $P_{f,3}$ while keeping the transverse momentum transfer and $z$ fixed changes more than one of $t$, $\xi$, and $\nu$ simultaneously. As a result, the GFFs cannot be fitted separately at each $t$, requiring a global fit to the entire dataset. Fitting the GFFs point by point in $t$ with a $z$-expansion~\cite{Boyd:1994tt,Hill:2010yb,Bourrely:2008za} or a dipole form~\cite{Hagler:2007xi,Alexandrou:2019ali}, as is commonly done, is therefore not possible here. Such forms could still be used in a global fit, but they require choosing a functional form and its truncation in advance. We instead employ an ANN-based parametrization. This procedure is a global fit, where all relevant kinematics and coupled GFFs are fitted simultaneously, and the network plays the role that a dipole or a $z$-expansion ansatz would play, with the difference that no analytical form is assumed and that the fit is performed directly at the individual values of $t$, without binning. As such, ANNs can appropriately accommodate all types of variations in $t$. The expression for $t$ given in Eq.~\eqref{eq:kinematics} can be further expanded in terms of temporal and spatial components,
\begin{align}
    t = \left(E(\vec{P}_f, m) - E(\vec{P}_i, m)\right)^2 - \vec{\Delta}^2.
    \label{eq:momentum-transfer-squared}
\end{align}
In the case of zero-skewness GPDs, where $P_{f,3} = P_{i,3} = P_3$, changing $\nu$ by changing $P_3$ for a fixed value of $\vec{\Delta}$ alters the value of $t$ only slightly in $\text{GeV}^2$, as shown in Table~\ref{tab:zero-skewness-t}. Earlier calculations often ignored these variations and approximated them using a single average value~\cite{Bhattacharya:2023ays}. Such an approximation is particularly inappropriate for nonzero skewness ($P_{f,3} \neq P_{i,3}$), where for the same transverse momentum transfer, changing $P_{3}$ corresponds to different $\xi$ and to values of $t$ that differ substantially, as shown in Table~\ref{tab:nonzero-skewness-t}. ANNs are well suited for this task, since they fit all kinematic data globally and can directly accommodate these variations in $t$.

\subsection{Architecture}
\label{sec:ann-arch}

\begin{figure}[tb]
\centering
\includegraphics[width=0.9\columnwidth]{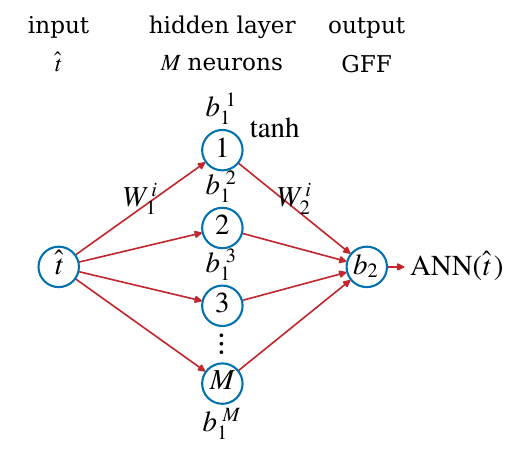}
\caption{The structure of the ANN used for each GFF.}
\label{fig:ann}
\end{figure}

We employ a simple feed-forward architecture, consisting of a one-dimensional input, one hidden layer containing $M$ neurons, and a one-dimensional output for each GFF,
\begin{align}
\mathrm{ANN}(\hat{t}) &= b_2 + \sum_{i=1}^{M} W_2^{\,i}\,\tanh\!\left( W_1^{\,i}\,\hat{t} + b_1^{\,i} \right) .
\label{eq:ann}
\end{align}
This network takes the rescaled momentum transfer squared $\hat{t}$ as an input, as defined in Eq.~\eqref{eq:rescale}. $W_1^{\,i}$ and $b_1^{\,i}$ are the weights and biases connecting the input layer to the hidden layer, and $W_2^{\,i}$ and $b_2$ are the corresponding parameters connecting the hidden layer to the output. The activation function is chosen as hyperbolic tangent. A network with a single hidden layer is a universal approximator \cite{Cybenko:1989iql,Hornik:1989yye}, so that the parametrization is flexible and does not impose an analytical functional form on the result. For this analysis we use $M=5$ neurons, which provides sufficient flexibility for the $t$ dependence while keeping the number of parameters moderate. Not all of these parameters are tightly constrained by the data, since the higher GFFs enter the fit only through strongly suppressed terms, as discussed in Sec.~\ref{sec:results-fs}. This freedom is reflected in the systematic uncertainty associated with the network architecture in Sec.~\ref{sec:sys-ann}. The architecture is illustrated in Fig.~\ref{fig:ann}. For our work, weights are initialized with uniform random numbers, $W_1^{\,i},W_2^{\,i}\sim\mathcal{U}(-1,1)$, while all biases are initialized to zero.

\begin{table}[t]
\caption{Hadron boost momenta $P_3$ and corresponding momentum transfer $-t$ at zero skewness for $\vec{\Delta} = (0.833, 0, 0)$~GeV.}
\label{tab:zero-skewness-t}
\begin{ruledtabular}
\begin{tabular}{ccc}
$n_{P_3}$ & $P_3$ [GeV] & $-t$ [GeV$^2$] \\
\colrule
$1$ & $0.417$ & $0.620$ \\
$2$ & $0.833$ & $0.638$ \\
$3$ & $1.250$ & $0.654$ \\
$4$ & $1.666$ & $0.666$
\end{tabular}
\end{ruledtabular}
\end{table}

\begin{table}[t]
\caption{Momentum transfer $-t$ at fixed sink momentum $n_{p,f}=4$ and fixed transverse momentum transfer $(\Delta_1,\Delta_2)=(0.833,0)$~GeV, for the initial-state momenta $P_{i,3}=n_{p,i}(2\pi/L)$ of the new data set. The row with $n_{p,i}=4$ is the zero-skewness case of Table~\ref{tab:zero-skewness-t}, shown for reference. The proton mass is the same as in Table~\ref{tab:zero-skewness-t}.}
\label{tab:nonzero-skewness-t}
\begin{ruledtabular}
\begin{tabular}{ccccc}
$n_{p,f}$ & $n_{p,i}$ & $\xi$ & $P_3$ [GeV] & $-t$ [GeV$^2$] \\
\colrule
$4$ & $2$ & $-1/3$ & $1.250$ & $1.236$ \\
$4$ & $3$ & $-1/7$ & $1.458$ & $0.849$ \\
$4$ & $4$ & $0$    & $1.666$ & $0.666$ \\
$4$ & $5$ & $1/9$  & $1.875$ & $0.612$ \\
$4$ & $6$ & $1/5$  & $2.083$ & $0.640$
\end{tabular}
\end{ruledtabular}
\end{table}

For reasons of training stability, the input of the network is not $t$ itself but a rescaled variable. Denoting by $t_{\rm min}$ the smallest $t$ entering the fit, we map the accessible interval $t\in[t_{\rm min},0]$ onto $[-1,1]$ using
\begin{align}
\hat{t} = 1-\frac{2t}{t_{\rm min}}.
\label{eq:rescale}
\end{align}
This is a technical rescaling which maps the fitted \(t\) interval onto \([-1,1]\), keeping the ANN inputs dimensionless and of order unity and improving the optimization. It also removes the dependence of the input scale on the size of the fitted \(t\) interval. In practice, it improves the convergence of the gradient-descent minimization described in Sec.~\ref{sec:ann-training}.

\subsection{Towards parametrization of GFFs}
\label{sec:ann-param}

As mentioned above, the GFFs such as $G\in\{A_{n+1,i},\,B_{n+1,i},\,C_{n+1}\}$ are represented in our calculations with $\text{ANN}(\hat{t})$,
\begin{align}
G(t) &= b_2^{\,G} + \sum_{i=1}^{M} W_2^{\,G,i}\,\tanh\!\left( W_1^{\,G,i}\,\hat{t} + b_1^{\,G,i} \right) ,
\label{eq:ann-gff}
\end{align}
with $\hat{t}$ being calculated as per Eq.~\eqref{eq:rescale}, and with an independent set of parameters $\theta_G\in\{W_1^{\,G,i},\,b_1^{\,G,i},\,W_2^{\,G,i},\,b_2^{\,G}\}$. For $M=5$, there are $16$ free parameters for each GFF, yielding $224$ parameters in total to be constrained by the lattice data. 

Such a parametrization of the GFFs remains completely unconstrained, assuming nothing beyond the polynomiality relations of Eqs.~\eqref{eq:polyH} and \eqref{eq:polyE}. In particular, we do not enforce the normalizations in the forward limit,
\begin{align}
A_{1,0}(0) &= 1 , \; B_{1,0}(0) = \kappa_u - \kappa_d ,
\label{eq:norm}
\end{align}
for the isovector combination, where $\kappa_{u}$ and $\kappa_{d}$ are the quark contributions to the anomalous magnetic moment of the proton, even though the first relation is fixed exactly by the isovector charge of the proton. This choice is motivated by two reasons. Firstly, the values of $\kappa_q$ are not inputs to a lattice calculation, so that imposing the second relation would introduce experimental information into a first-principles determination. Secondly, leaving $A_{1,0}(0)$ free provides a nontrivial consistency check of the framework, since its value at $t=0$ must then be reproduced by the data alone. 

The model prediction for the reduced ITDs is linear in the GFFs, as follows from the polynomiality relations of Eqs.~\eqref{eq:polyH}--\eqref{eq:polyE} combined with the short-distance relation of Eq.~\eqref{eq:matching-moments}. Explicitly, inserting Eqs.~\eqref{eq:mom1}--\eqref{eq:mom4} into Eqs.~\eqref{eq:ReF} and \eqref{eq:ImF},
\begin{align}
\mathrm{Re}\,F^{H} &= K_1\,A_{1,0} + K_3\left( A_{3,0} + \xi^2 A_{3,2} \right) , \label{eq:designReH} \\
\mathrm{Im}\,F^{H} &= K_2\left( A_{2,0} + \xi^2 C_2 \right) \nonumber\\
&+ K_4\left( A_{4,0} + \xi^2 A_{4,2} + \xi^4 C_4 \right) , \label{eq:designImH} \\
\mathrm{Re}\,F^{E} &= K_1\,B_{1,0} + K_3\left( B_{3,0} + \xi^2 B_{3,2} \right) , \label{eq:designReE} \\
\mathrm{Im}\,F^{E} &= K_2\left( B_{2,0} - \xi^2 C_2 \right) \nonumber\\
&+ K_4\left( B_{4,0} + \xi^2 B_{4,2} - \xi^4 C_4 \right) , \label{eq:designImE}
\end{align}
where the kinematic factors $K_{1,2,3,4}$ collect the Ioffe-time prefactors and the Wilson coefficients of Eqs.~\eqref{eq:ReF}--\eqref{eq:ImF},
\begin{align}
K_1 &= 1 - \frac{\nu^2}{2}\,\xi^2 c_{3,2} ,\; K_3 = -\frac{\nu^2}{2}\,c_{3,0},\nonumber\\
K_2 &= -\nu\left( c_{2,0} - \frac{\nu^2}{6}\,\xi^2 c_{4,2} \right) ,\; K_4 = \frac{\nu^3}{6}\,c_{4,0}.
\label{eq:Kfactors}
\end{align}
Collecting the fourteen GFFs into the vector
$\mathbf{G}(\hat{t})=(A_{1,0},\ldots,C_4)^T$, we label each predicted reduced ITD by an index $i$, which specifies the GPD type ($H$ or $E$),
the real or imaginary component, and the kinematics
$(z,P_{f,3},\vec{\Delta})$. The latter determine $\nu$, $\xi$, and $t$.
The corresponding prediction can then be written as
\begin{align}
F_i^{\rm pred}
&=
\sum_{k=1}^{14}
X_{ik}(z,P_{f,3},\vec{\Delta})\,
G_k\!\left(\hat{t}_i;\theta_k\right),
\label{eq:design}
\end{align}
where $G_k(\hat{t}_i;\theta_k)$ denotes the ANN parametrization of the
$k$th GFF and the coefficients $X_{ik}$ are read off from
Eqs.~\eqref{eq:designReH}--\eqref{eq:designImE}. The prediction for the whole data set is thus a single matrix product, which can be evaluated for all kinematic points at once.

Equations~\eqref{eq:designReH}--\eqref{eq:designImE} also show explicitly which data constrain which GFFs. The real part constrains $A_{1,0}$, $A_{3,0}$, $A_{3,2}$ and their $B$-type counterparts. The imaginary part constrains $A_{2,0}$, $A_{4,0}$, $A_{4,2}$, $B_{2,0}$, $B_{4,0}$ and $B_{4,2}$, alongside $C_2$ and $C_4$. The last two GFFs appear only in the combinations $A_{2,0}+\xi^2 C_2$ and $B_{2,0}-\xi^2 C_2$, and similarly for $C_4$. Since $C_2$ and $C_4$ are common to both channels, we determine them from a simultaneous fit to the $H$- and $E$-type data, which uses the most information. Determinations from a single channel are also possible and are used in Sec.~\ref{sec:sys-dterm} as a test of this choice.

\subsection{Training and estimation of statistical uncertainty}
\label{sec:ann-training}

The parameters of the network are determined by minimizing a $\chi^2$, also referred to as the loss function, that quantifies the agreement between the model predictions and the lattice data for the reduced ITDs. The real and the imaginary parts are normalized separately, so that the convergence of the two can be monitored independently. This is useful because the two parts constrain disjoint sets of GFFs and are affected differently by the systematic effects:
\begin{align}
\chi^2 &= \frac{1}{N_{\rm Re}}\sum_{i\in\mathrm{Re}} \left[ \frac{F_i^{\rm pred}(\theta) - F_i^{\rm lat}}{\sigma_i^{\rm lat}} \right]^2 \nonumber\\
&+ \frac{1}{N_{\rm Im}}\sum_{i\in\mathrm{Im}} \left[ \frac{F_i^{\rm pred}(\theta) - F_i^{\rm lat}}{\sigma_i^{\rm lat}} \right]^2 ,
\label{eq:chi2}
\end{align}
where the sums run over all included kinematic points $
i \equiv \bigl(\vec{\Delta},P_{f,3},z\bigr)
$ with both \(H\) and \(E\) data entering the same loss, and $N_{\rm Re}=N_{\rm Im}$ for the data sets used here. The training set comprises the nonforward reduced ITDs at $\xi=0$ and $\xi\neq0$ together with the forward points at $t=\xi=0$. The minimization with respect to the network parameters $\theta$ is
performed using full-batch gradient descent with a fixed learning rate.
The gradients of the loss function with respect to $\theta$ are evaluated
analytically by applying the chain rule through
Eqs.~\eqref{eq:chi2}, \eqref{eq:design}, and \eqref{eq:ann-gff}. This simple optimization prescription is sufficient for the relatively small networks employed here. 

Among the usual criteria for a fit, the most relevant one here is the convergence of the $\chi^2$ during the minimization. $\chi^2$ converges to a stable value if the parametrization is sufficiently constrained by the available data. This convergence is tested by varying the number of neurons, the initialization of the parameters and the learning rate within the same architecture, and more details can be found in Sec.~\ref{sec:sys-ann}.
 Training quality is also inspected through direct comparison between the reconstructed observables and the lattice data shown in Sec.~\ref{sec:results-fs}.

The statistical uncertainties are calculated using bootstrap resampling. The lattice-calculated matrix elements are evaluated on a set of 101 bootstrap samples, constructed by resampling the gauge configuration numbers with replacement, and the networks are trained on each sample starting from the same initialization of the parameters, preventing initialization dependence from entering the statistical uncertainty. For our analysis, central values are always determined from sample 0, which corresponds to the un-resampled configuration set where each configuration number appears exactly once. Hence, we estimate the central values of the GFFs by training the associated networks on sample 0, while we obtain the statistical uncertainties by training the networks repeatedly on each bootstrap sample and then taking the standard deviation over the samples.
The stochasticity of the central value is studied in Sec.~\ref{sec:sys-ann}, where we test the effects of different initializations. 

\subsection{Summary of the procedure}
\label{sec:ann-summary}
To summarize, the fourteen GFFs are parametrized by independent neural networks, the reduced ITDs are predicted from them through Eqs.~\eqref{eq:designReH}--\eqref{eq:designImE}, and the network parameters are fitted to the lattice data on every bootstrap sample to estimate the statistical uncertainty. In Sec.~\ref{sec:results}, we define the \(z\)- and \(t\)-range used to select the lattice data entering into the fit, together with the truncation order and network hyperparameters, and assess the impact of these choices in Sec.~\ref{sec:systematics}.

\section{Numerical Results }
\label{sec:results}
\subsection{Lattice setup}
\label{sec:lattice}
We use the bare coordinate-space matrix elements of Ref.~\cite{Chu:2025kew} at the three fixed sink momenta $P_{f,3}=n_{p,f}(2\pi/L)$, with $n_{p,f}=\pm1,\pm2,\pm3$, and add a new data set at $n_{p,f}=\pm4$. The calculations use an $N_f=2+1+1$ ensemble of clover-improved Wilson twisted-mass fermions and the Iwasaki-improved gauge action~\cite{Alexandrou:2018egz}, with a lattice volume of $32^3\times64$, a lattice spacing $a\simeq0.093$~fm and a pion mass $m_\pi\simeq260$~MeV. The computation employs randomly distributed source positions on each gauge configuration, with each source serving as the origin for a Dirac operator inversion, in order to enhance the statistical precision. We study the flavor-nonsinglet $u-d$ channel and use a source-sink separation $t_s=10a\simeq0.93$~fm for all four sink momenta. Table~\ref{tab:lattice-statistics} summarizes the lattice statistics. The forward matrix elements at $t=\xi=0$ are computed on the same ensembles and are taken from the same reference. A more detailed description of the kinematic setup, including the momentum smearing parameters, is given in Ref.~\cite{Chu:2025kew}.

As a reminder, $P_{f,3}$ is the fixed sink momentum, while $P_3=(P_{i,3}+P_{f,3})/2$ is the average longitudinal momentum entering the Ioffe time $\nu=P_3z$. For the new $n_{p,f}=4$ data, the initial-state momenta $P_{i,3}=n_{p,i}(2\pi/L)$ with $n_{p,i}=2,3,5,6$ give $(n_{p,i},\xi)=(2,-1/3)$, $(3,-1/7)$, $(5,1/9)$ and $(6,1/5)$, where
\begin{align}
\xi=-\frac{\Delta_3}{2P_3}
=\frac{P_{i,3}-P_{f,3}}{P_{i,3}+P_{f,3}}.
\end{align}
At each value of $\xi$, we include the transverse momentum-transfer classes $(|\Delta_1|,|\Delta_2|)=(1,0),(1,1),(1,2),(1,3),(2,0),(2,2),(3,0),(4,0)$, in units of $2\pi/L$. These new data reach an average longitudinal momentum of $P_3\simeq2.08$~GeV  and therefore extend the range of Ioffe times entering the fit, which is relevant for the truncation of Eq.~\eqref{eq:nu-expansion}.

\begin{table}[t]
\caption{Lattice statistics for each fixed sink momentum $P_{f,3}=n_{p,f}(2\pi/L)$. Here, $N_{\rm src}$ is the number of source positions per gauge configuration. The $n_{p,f}=1,2,3$ data are from Ref.~\cite{Chu:2025kew}, while the $n_{p,f}=4$ data are new to this work.}
\label{tab:lattice-statistics}
\begin{ruledtabular}
\begin{tabular}{ccccc}
$n_{p,f}$ & $|P_{f,3}|$ [GeV] & $N_{\rm cfg}$ & $N_{\rm src}$ & $t_s/a$ \\
\colrule
$1$ & $0.42$ & $96$  & $8$  & $10$ \\
$2$ & $0.83$ & $95$  & $8$  & $10$ \\
$3$ & $1.25$ & $120$ & $32$ & $10$ \\
$4$ & $1.67$ & $345$ & $32$ & $10$
\end{tabular}
\end{ruledtabular}
\end{table}

\begin{figure}[tb]
\centering
\includegraphics[width=\columnwidth]{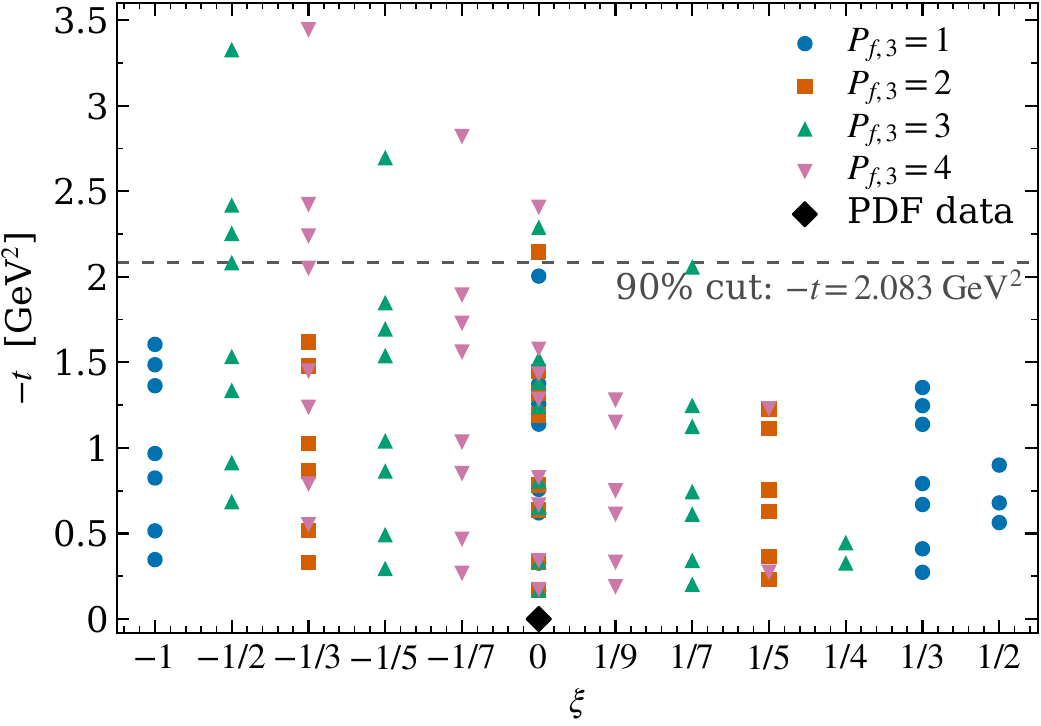}
\caption{Kinematic coverage of the lattice data in the $(\xi,-t)$ plane. The data comprise $110$ values of $t$, of which $99$ lie below the horizontal dashed line marking $-t=2.083~{\rm GeV}^2$. These are the points retained in the central analysis. Circles, squares, upward triangles, and downward triangles denote the fixed sink momenta $P_{f,3}=n_{p,f}(2\pi/L)$ with $n_{p,f}=1,2,3,4$, respectively. The $n_{p,f}=1,2,3$ data are from Ref.~\cite{Chu:2025kew}, while the $n_{p,f}=4$ data are new to this work. The black diamond marks the common forward point at $(\xi,t)=(0,0)$ for all four momenta.}
\label{fig:kinematic-coverage}
\end{figure}

Figure~\ref{fig:kinematic-coverage} shows the combined kinematic coverage of the nonforward GPD matrix elements and of the forward PDF matrix elements. The four sink momenta, combined with the initial-state momenta listed above, give access to both positive and negative skewness in the range $-1\leq\xi\leq1/2$ and to $110$ values of the momentum transfer. The forward data at $(\xi,t)=(0,0)$ directly constrain $A_{n+1,0}(0)$, the forward limits of the $A$-type GFFs. At sufficiently large $z$, the
perturbative treatment of the matching coefficients becomes unreliable. Moreover, as discussed in Sec.~\ref{sec:sde}, the short-distance relation of Eq.~\eqref{eq:matching-moments} receives power corrections of order $\mathcal{O}(z^2\Lambda_{\rm QCD}^2)$, $\mathcal{O}(z^2m^2)$ and $\mathcal{O}(z^2t)$. To reduce these corrections and improve convergence of perturbation theory, we restrict the central analysis to short Wilson lines, $z/a=2,3$, and to $-t\leq2.08~{\rm GeV}^{2}$, the latter cut discarding the $10\%$ of the nonforward points with the largest $|t|$, which are also the noisiest, while $z/a=1$ is discarded because of possible enhanced discretization effects. The same $z$ window and $t$ cut are used in the fixed energy scale and in the RGR analyses, so that the two can be compared directly. The dependence on the $z$ window and the  $t$ cut is studied in Secs.~\ref{sec:sys-z} and~\ref{sec:sys-tcut}, respectively.

\subsection{Results on Fixed Energy Scale Prescription}
\label{sec:results-fs}

We first present the results obtained with the fixed energy scale prescription of Sec.~\ref{sec:sde}, with all Wilson coefficients evaluated at NLO in the $\overline{\rm MS}$ scheme at the reference scale $\mu=2$~GeV. In our standard analysis, we use the previously mentioned $z$- and $t$-cuts. $C_2$ and $C_4$ are determined from a simultaneous fit to the $H$- and $E$-type channels. Unless stated otherwise, the results below refer to this standard setup.

\begin{figure}[t]
\centering
\includegraphics[width=\columnwidth]{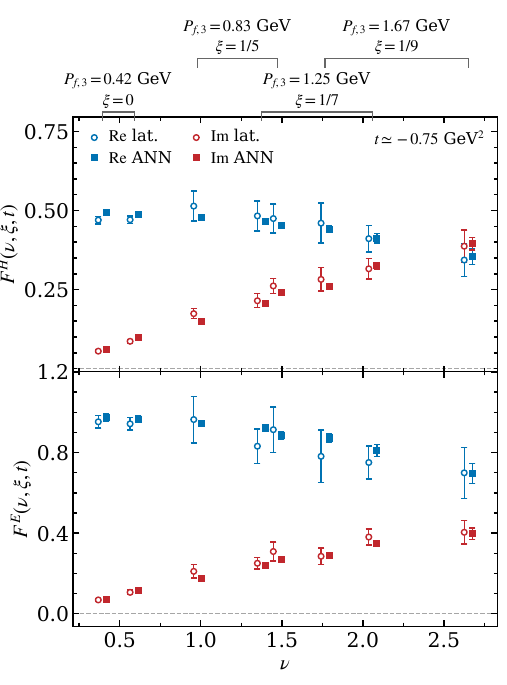}
\caption{Comparison between the lattice reduced ITDs and their reconstruction from the ANN extraction in the fixed-scale analysis at $\mu=2$~GeV, at $t\simeq-0.75$~GeV$^2$. Open circles denote the lattice data and filled squares the ANN result. Both are shown with their statistical uncertainties and are displaced horizontally by $\pm0.025$ in $\nu$ for legibility, although each pair corresponds to the same Ioffe time. The upper panel shows the real (blue) and imaginary (red) parts of the $H$-type reduced ITD and the lower panel the same for the $E$-type one. Results are shown for the four sink momenta $P_{f,3}=0.42$, $0.83$, $1.25$ and $1.67$~GeV with $\xi=0$, $1/5$, $1/7$ and $1/9$, respectively, indicated by the brackets. The two points within each bracket correspond to $z/a=2$ and $3$. The individual momenta sample $t$ between $-0.757$ and $-0.745$~GeV$^2$. All points shown enter the ANN training.}
\label{fig:itd-fs-t075}
\end{figure}

Fig.~\ref{fig:itd-fs-t075} compares the lattice-reduced ITDs with those reconstructed from the fitted GFFs using Eqs.~\eqref{eq:designReH}--\eqref{eq:designImE}. The figure collects the four sink momenta at an approximately common momentum transfer, $t\simeq-0.75$~GeV$^2$, so that the momentum and skewness dependence can be examined at fixed $t$. Each momentum is paired with a different skewness, so the four brackets also scan $\xi$ from $0$ to $1/5$, and the two entries within each bracket correspond to the two Wilson line lengths $z/a=2$ and $3$. The reconstruction agrees with the lattice data at every point shown in the plot for both the real and imaginary parts of the $H$- and $E$-type channels, within one standard deviation for $25$ of the $32$ entries and within two standard deviations for all of them. The differences in prediction show no systematic trend with $\nu$, momentum, or skewness. The same comparison has been carried out at the remaining values of the momentum transfer, covering the full skewness range of the data set, $-1\leq\xi\leq1/2$. The level of agreement is consistent throughout, although it is naturally harder to assess where the data are noisy or sparse. Those figures are not reproduced here. For one of these kinematics, Appendix~\ref{app:hyper} compares the predictions of the fit with the lattice data at all available Wilson line lengths, including those that are not used in the training.

The uncertainties of the reconstructed ITDs are visibly smaller than those of the lattice points at most kinematics, which is the expected behavior rather than an indication of an underestimated error: the uncertainty shown for the reconstruction is that of a global fit propagated to a single kinematic point, and not a prediction of the error bar of an individual measurement. The $808$ real and imaginary entries that enter the standard fit, namely $792$ nonforward entries of $F^H$ and $F^E$ and $16$ forward ones, constrain fourteen functions of $t$, so that information from many kinematics is combined at every point of the reconstruction. How much the reconstructed uncertainty falls below the lattice one varies across the data set, because the statistical precision of the lattice matrix elements differs widely from one kinematic point to another. The quoted errors span three orders of magnitude, from $\sigma\simeq0.002$ to $\sigma\simeq1.8$, so that the $10\%$ most precise entries carry $67\%$ of the total $\chi^2$ weight. The fit is therefore anchored to the high-precision subset, namely the zero-skewness matrix elements at small $|t|$ and $z/a=2$, while the least precise entries, which lie mostly at $\xi=1/5$ and $1/4$ in the $E$-type channel, have only a limited influence on the minimization. This behavior is directly visible in Fig.~\ref{fig:itd-fs-t075}. At $P_{f,3}=0.42$~GeV and $\xi=0$, the lattice and reconstructed uncertainties are comparable, differing by a factor between $1.5$ and $2.9$, whereas at $P_{f,3}=1.67$~GeV, the lattice uncertainty exceeds the reconstructed one by up to a factor of about seven, the description at that momentum being inherited from the rest of the data set. Where the lattice data are already precise, the reconstructed uncertainty does not fall appreciably below them, which supports the consistency of the statistical uncertainty propagation. The agreement between the ANN predictions and the lattice reduced-ITD data across the fitted kinematic range shows that the extracted GFFs consistently describe the data.

\begin{figure*}[tp]
\centering
\includegraphics[width=0.245\textwidth]{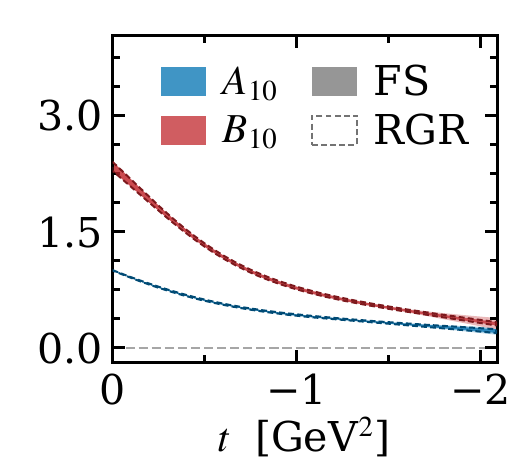}
\includegraphics[width=0.245\textwidth]{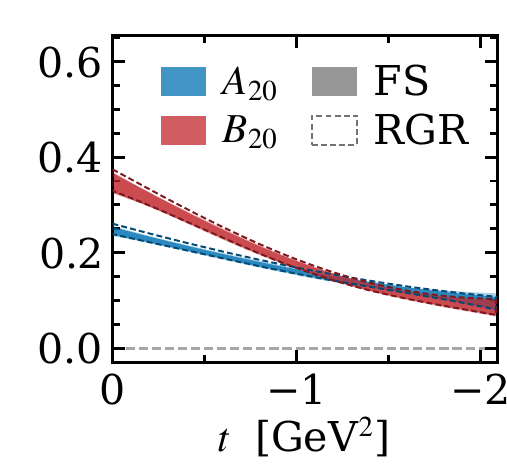}
\includegraphics[width=0.245\textwidth]{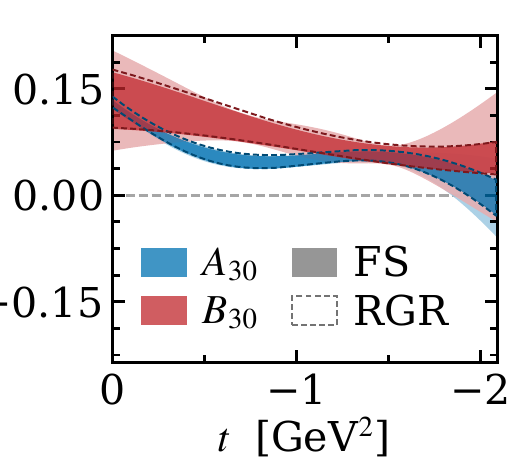}
\includegraphics[width=0.245\textwidth]{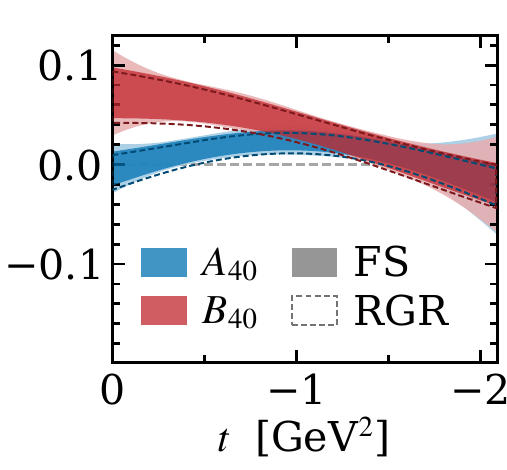}
\includegraphics[width=0.245\textwidth]{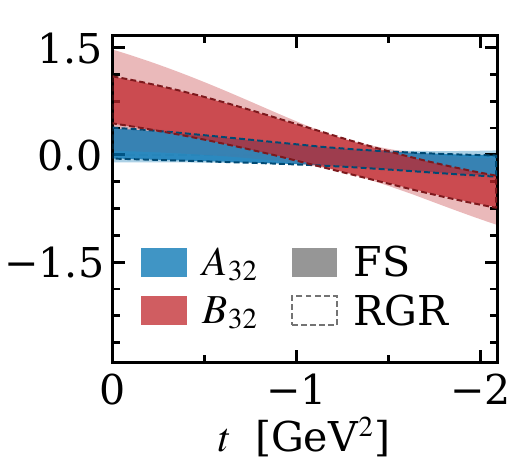}
\includegraphics[width=0.245\textwidth]{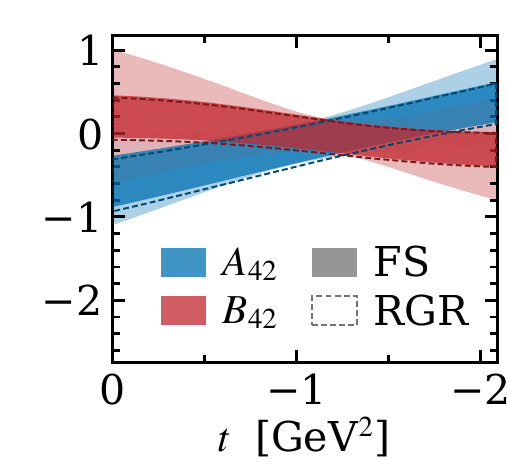}
\includegraphics[width=0.245\textwidth]{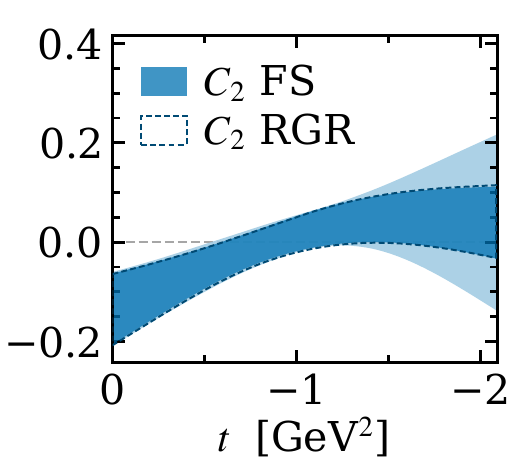}
\includegraphics[width=0.245\textwidth]{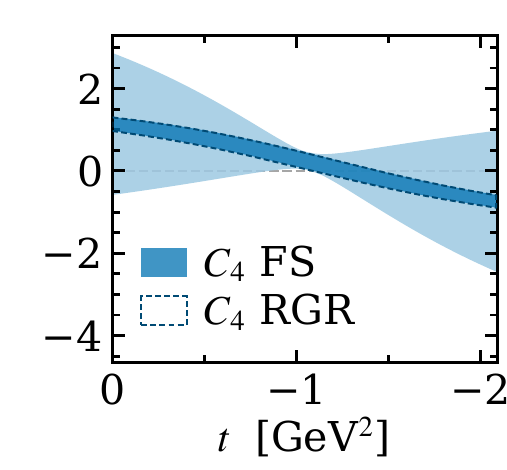}
\caption{Generalized form factors at $\mu=2$~GeV in the $\overline{\rm MS}$ scheme and at NLO, from the fixed energy scale (FS) and the RGR analyses. The panels show, in reading order, $(A_{1,0},B_{1,0})$, $(A_{2,0},B_{2,0})$, $(A_{3,0},B_{3,0})$, $(A_{4,0},B_{4,0})$, $(A_{3,2},B_{3,2})$, $(A_{4,2},B_{4,2})$, $C_2$, and $C_4$. The filled bands show the baseline FS analysis: the inner bands denote the statistical uncertainties propagated through the ANN extraction and the outer bands the total uncertainties, obtained by adding the systematic uncertainties of Sec.~\ref{sec:systematics} and the statistical uncertainties in quadrature. The dashed frames show the RGR analysis with $\kappa=\sqrt{2}$ and delimit its total uncertainty, obtained by adding in quadrature the statistical uncertainty and the systematic uncertainty from the variation of $\kappa$ described in Sec.~\ref{sec:results-rgr}.}
\label{fig:gff-fs}
\end{figure*}

The extraction of the GFFs depends on several choices of analysis, which we vary in Sec.~\ref{sec:systematics}. For each source of systematic uncertainty, the corresponding uncertainty is estimated as described in Sec.~\ref{sec:systematics}. The individual systematic uncertainties and the statistical uncertainty are then added in quadrature to obtain the total uncertainty. In Fig.~\ref{fig:gff-fs}, the inner filled band denotes the statistical uncertainty of the FS analysis and the outer filled band its total uncertainty, while the dashed frame denotes the total uncertainty of the RGR analysis discussed in Sec.~\ref{sec:results-rgr}.

We present the GFFs obtained from the ANN fit in Fig.~\ref{fig:gff-fs}. The $A$- and $B$-type GFFs are grouped according to the order of the Mellin moment in which they appear, while $C_2$ and $C_4$ are shown separately. The results display a clear hierarchy in the quality of the constraints. The GFFs with $i=0$, namely $A_{n+1,0}$ and $B_{n+1,0}$, are in general well determined and have a signal significantly different from zero over a sizable range of $t$, with the exception of $A_{4,0}$, which is compatible with zero within its total uncertainty over a large part of the fitted range. In contrast, the GFFs that enter only through the nonzero-skewness terms, $A_{3,2}$, $B_{3,2}$, $A_{4,2}$ and $B_{4,2}$, are largely consistent with zero within the present precision. $C_2$ is weakly constrained, with uncertainties still too large to support a firm conclusion about its $t$ dependence, while $C_4$ has a very large systematic uncertainty, indicating that it is poorly constrained by the fit. The systematic uncertainty is comparable to or larger than the statistical one for most of the GFFs of the third and fourth moments, and exceeds it substantially for $B_{3,2}$, $B_{4,2}$ and $C_4$, where the outer band is visibly wider than the inner one.

This hierarchy is not an artifact of the parametrization, which assumes no functional form, but follows from the structure of Eqs.~\eqref{eq:designReH}--\eqref{eq:designImE} together with the kinematic coverage of the present data. The contributions of the higher Mellin moments enter with increasing powers of $\nu^{n}/n!$, while the GFFs associated with the skewness receive an additional suppression by $\xi^{2}$ or $\xi^{4}$, and the accessible values of $|\xi|$ are small for most of the data. In addition, the PDF data constrain $A_{n+1,0}(t)$ at $t=0$, so that the forward limits of $A_{1,0}$, $A_{2,0}$, $A_{3,0}$ and $A_{4,0}$ are appropriately determined by the combined fit, whereas no corresponding forward information is available for the $B$-type GFFs, whose extrapolation towards $t=0$ consequently carries larger uncertainties. A more precise determination will require a substantially denser coverage of the relevant kinematic variables. In particular, a larger physical volume would provide a finer momentum resolution and hence a denser sampling of the small-$|t|$ region. This is especially important in the present calculation, where there is a sizable gap between $t=0$ and the smallest nonzero momentum transfer. A finer lattice spacing would additionally provide more lattice separations within the same physical short-distance window, allowing the $z$ dependence to be sampled more densely than with the presently used $z/a=2,3$, while also reducing discretization effects. The present calculation should therefore be regarded primarily as a proof of concept. Increasing the kinematic coverage, improving the resolution near $t=0$, and employing finer lattice spacings provide a clear path toward a precision determination of the GFFs.

\subsection{Results on Renormalization Group Resummation Prescription}
\label{sec:results-rgr}

As described in Sec.~\ref{sec:sde}, in the FS prescription, the Wilson coefficients are evaluated at $\mu=2$~GeV, so the logarithm $L=\ln\left(e^{2\gamma_E+1}z^2\mu^2/4\right)$ in Table~\ref{tab:wilson} may not be small for the separations entering the fit, therefore potentially breaking down the convergence of perturbation theory used for computing these coefficients. The RGR prescription removes the $z$ dependence of this logarithm by evaluating the coefficients at the scale set by the separation itself, $\mu_0=2\kappa e^{-\gamma_E}/z$, and evolving the result to $\mu=2$~GeV, as described by Eq.~\eqref{eq:rgr}, resulting in the Wilson coefficients of Eqs.~\eqref{eq:crgr-diag}--\eqref{eq:crgr-42}. The RGR extraction introduces one additional parameter, the constant $\kappa$ that fixes the matching scale $\mu_{0}$. Larger values of $\kappa$ move $\mu_0$ into a region where the coupling is smaller, and the perturbative expansion is better behaved, at the price of a larger logarithm $l$, while smaller values of $\kappa$ do the opposite. We have repeated the analysis with $\kappa=1/\sqrt{2}$, $1$, $\sqrt{2}$, $2$ and $2\sqrt2$, and the dependence of the GFFs on $\kappa$ is discussed in Sec.~\ref{sec:sys-kappa}. The baseline RGR analysis uses $\kappa=\sqrt{2}$. Since the formalism does not fix \(\kappa\), the residual dependence of the GFFs on this parameter is a finite-order perturbative truncation effect. In an all-orders treatment, the result at \(\mu=2\) GeV would be independent of the intermediate scale \(\mu_0\), and hence of \(\kappa\). We therefore treat it as a source of systematic uncertainty. We estimate it as the average of the absolute differences between the results at $\kappa=1$ and $\kappa=2$ and the baseline $\kappa=\sqrt{2}$, which vary the matching scale by a factor $\sqrt{2}$ in either direction, and add it in quadrature to the statistical uncertainty.

We repeat the GFF extraction, keeping every other ingredient identical to the baseline analysis used in the FS prescription. The RGR results are shown in Fig.~\ref{fig:gff-fs} as dashed frames on top of the FS bands, and Table~\ref{tab:fs-rgr} compares the resulting GFFs at $t=0$ with those of the FS analysis, together with the largest difference between the two prescriptions over the whole range $-2.08\leq t\leq0$~GeV$^2$, in percent of the FS total uncertainty. Overall, the differences are small. The largest ones are found for $A_{2,0}$ and $A_{3,0}$, both with $51\%$ of the total uncertainty, followed by $B_{2,0}$ and $A_{4,0}$ with $32\%$ and $26\%$, respectively, while for all other GFFs they stay below $20\%$. The RGR central values therefore lie inside the FS total band over the full range of $t$, and the RGR frames overlap the FS bands in every panel of Fig.~\ref{fig:gff-fs}. The $\kappa$ uncertainty is sizable only for $A_{2,0}$, $A_{3,0}$ and $B_{2,0}$, where at $t=0$ it reaches $83\%$, $63\%$ and $44\%$ of the statistical uncertainty, respectively, while it is negligible for $A_{1,0}$ and $B_{1,0}$ and for the weakly constrained GFFs, whose uncertainty is dominated by statistics.

These results match expectations. The real part of the reduced ITD is dominated by $A_{1,0}$ and $B_{1,0}$, which have vanishing anomalous dimensions and therefore do not undergo intrinsic scale evolution. Consequently, the real part is expected to be less sensitive to RGR, apart from the residual dependence entering through the matching and mixing with higher moments. Correspondingly, the real part of the loss is almost unchanged, $\chi^{2}_{\rm Re}/N=1.024$ in the fixed energy scale analysis and $1.020$ in the RGR one, while the imaginary part, which only carries the scale-dependent GFFs and is therefore more directly sensitive to the RGR treatment, slightly improves from $\chi^{2}_{\rm Im}/N=1.293$ to $1.234$. Since the extraction is performed through a global fit, the higher moments can indirectly influence \(A_{1,0}\) and \(B_{1,0}\), but Table~\ref{tab:fs-rgr} shows that this effect stays below $10\%$ of the total uncertainty, and the same reason makes their $\kappa$ uncertainty negligible. The largest differences appear instead for $A_{2,0}$ and $A_{3,0}$, followed by $B_{2,0}$, which are the most precisely determined of the scale-dependent GFFs and enter with the scale-dependent Wilson coefficients $c_{2,0}$ and $c_{3,0}$. These are also the GFFs for which the $\kappa$ dependence is most visible (see Fig.~\ref{fig:sys-kappa-a20}).

Although the RGR prescription resums the logarithms associated with the Wilson-line separation, the data do not show a significant change in the extracted GFFs relative to the FS analysis. The two prescriptions give compatible results over the full fitted range in $t$. The RGR prescription introduces the additional auxiliary scale $\mu_0=2\kappa e^{-\gamma_E}/z$, and at finite perturbative order the extracted GFFs retain a residual dependence on $\kappa$, which for $A_{2,0}$ is comparable to the statistical uncertainty. We therefore adopt the FS analysis at $\mu=2$ GeV as our central result and use the RGR analysis, including its $\kappa$ uncertainty, as a complementary test of the perturbative matching.

\begin{table}[t]
\caption{Comparison of the GFFs at $t=0$ from the fixed energy scale and RGR analysis with $\kappa=\sqrt{2}$, both at $\mu=2$~GeV in the $\overline{\rm MS}$ scheme. In both columns the first uncertainty is statistical and the second systematic. For FS, the systematic uncertainty is that of Sec.~\ref{sec:sys-summary}. For RGR, it is the average of the absolute differences between the results at $\kappa=1$ and $\kappa=2$ and the baseline $\kappa=\sqrt{2}$. The last column gives the largest difference between the two central values over the fitted range $-2.08\leq t\leq0$~GeV$^2$, in percent of the FS total uncertainty, $\Delta_{\rm max}=\max_t|G^{\rm RGR}(t)-G^{\rm FS}(t)|/\sigma^{\rm FS}_{\rm tot}(t)$, where $\sigma^{\rm FS}_{\rm tot}$ includes the systematic uncertainties.}
\label{tab:fs-rgr}
\begin{ruledtabular}
\begin{tabular}{lccc}
GFF & FS & RGR & $\Delta_{\rm max}$ [\%] \\
\colrule
$A_{1,0}$ & $0.9995(29)(11)$ & $0.9998(29)(1)$ & $8$ \\
$B_{1,0}$ & $2.342(58)(28)$ & $2.342(58)(<1)$ & $3$ \\
$A_{2,0}$ & $0.2456(82)(65)$ & $0.2499(84)(69)$ & $51$ \\
$B_{2,0}$ & $0.347(20)(10)$ & $0.352(21)(9)$ & $32$ \\
$A_{3,0}$ & $0.1280(62)(33)$ & $0.1315(64)(40)$ & $51$ \\
$B_{3,0}$ & $0.133(40)(58)$ & $0.136(41)(3)$ & $15$ \\
$A_{4,0}$ & $-0.004(17)(18)$ & $-0.008(17)(3)$ & $26$ \\
$B_{4,0}$ & $0.072(25)(34)$ & $0.068(26)(3)$ & $19$ \\
$A_{3,2}$ & $0.16(21)(16)$ & $0.16(22)(<1)$ & $3$ \\
$B_{3,2}$ & $0.76(33)(63)$ & $0.77(33)(<1)$ & $1$ \\
$A_{4,2}$ & $-0.57(31)(43)$ & $-0.62(31)(4)$ & $15$ \\
$B_{4,2}$ & $0.20(25)(76)$ & $0.18(25)(2)$ & $10$ \\
$C_{2}$ & $-0.134(71)(22)$ & $-0.136(72)(2)$ & $5$ \\
$C_{4}$ & $1.14(17)(1.72)$ & $1.13(17)(1)$ & $4$ \\
\end{tabular}
\end{ruledtabular}
\end{table}

\subsection{Angular momentum and impact-parameter distributions}
\label{sec:pheno}

The GFFs extracted above give access to two physical quantities. The first is the quark total angular momentum of Eq.~\eqref{eq:jisumrule}, which follows from the forward limits $A_{2,0}(0)$ and $B_{2,0}(0)$ alone. We quote it from the fixed energy scale analysis. We obtain
\begin{align}
J^{u-d} = 0.296(11)_{\rm sta}(7)_{\rm sys} ,
\label{eq:jresult}
\end{align}
at $\mu=2$~GeV, where the first uncertainty is statistical and the second is the systematic uncertainty of Sec.~\ref{sec:sys-summary}. Table~\ref{tab:jsumrule} compares this value with selected previous lattice determinations from local~\cite{Hackett:2023rif,Alexandrou:2026oks} and nonlocal~\cite{Bhattacharya:2023ays,HadStruc:2024rix,Chu:2025jsi} operators. Our result agrees with the nonlocal determinations, with a total uncertainty about a factor of two smaller than that of Ref.~\cite{Chu:2025jsi}, which uses the same framework with zero-skewness data only. The local-operator results lie lower. Since the calculations use different ensembles, none is extrapolated to the continuum and physical-pion-mass limits, and excited-state effects are not perfectly controlled in the present work, this difference can be investigated in the future.

\begin{table}[t]
\caption{Quark total angular momentum of the isovector combination at $\mu=2$~GeV, compared with previous lattice determinations. The upper (lower) group of entries is obtained from matrix elements of local (nonlocal) operators. For the present work the first uncertainty is statistical and the second systematic, while for the other entries the uncertainty is the one quoted in the corresponding reference.}
\label{tab:jsumrule}
\begin{ruledtabular}
\begin{tabular}{lc}
& $J^{u-d}$ \\
\colrule
Hackett et al. (2024)~\cite{Hackett:2023rif}  & $0.2016(86)$ \\
ETMC (2026)~\cite{Alexandrou:2026oks}         & $0.214(36)$  \\
\colrule
ETMC (2023)~\cite{Bhattacharya:2023ays}       & $0.281(32)$  \\
HadStruc (2024)~\cite{HadStruc:2024rix}       & $0.288(21)$  \\
ANN reconstruction (2026)~\cite{Chu:2025jsi}  & $0.295(30)$  \\
This work                                     & $0.296(11)(7)$
\end{tabular}
\end{ruledtabular}
\end{table}

The second is the tomography of the proton. At zero skewness, the spatial momentum transfer is purely transverse, so that in the symmetric frame $t=-\vec{\Delta}_\perp^{\,2}$, and the GFFs with $i=0$ Fourier transform into impact-parameter distributions~\cite{Burkardt:2002hr,Diehl:2005jf},
\begin{align}
\rho_{n+1}(\vec{b}_\perp) &= \int\frac{{\rm d}^2\vec{\Delta}_\perp}{(2\pi)^2}\,A_{n+1,0}(-\vec{\Delta}_\perp^{\,2})\,e^{-i\vec{b}_\perp\cdot\vec{\Delta}_\perp},
\label{eq:density-unpol}\\
\rho^{T}_{n+1}(\vec{b}_\perp) &= \int\frac{{\rm d}^2\vec{\Delta}_\perp}{(2\pi)^2}\left[A_{n+1,0}(-\vec{\Delta}_\perp^{\,2})\right.\nonumber\\
&\qquad\left.+\,i\frac{\Delta_y}{2m}B_{n+1,0}(-\vec{\Delta}_\perp^{\,2})\right]e^{-i\vec{b}_\perp\cdot\vec{\Delta}_\perp},
\label{eq:density-pol}
\end{align}
where $m$ is the proton mass, $\rho_{n+1}$ is the distribution in an unpolarized proton and $\rho^{T}_{n+1}$ the one in a proton transversely polarized along $\hat{x}$, for which the $B$-type term produces a shift along $\hat{y}$. For $n=0$, these are the familiar charge and anomalous-magnetization distributions, while $n=1$ and $n=2$ weigh the same quantities by the momentum fraction $x$ and by $x^2$.

\begin{figure}[t]
\centering
\includegraphics[width=\columnwidth]{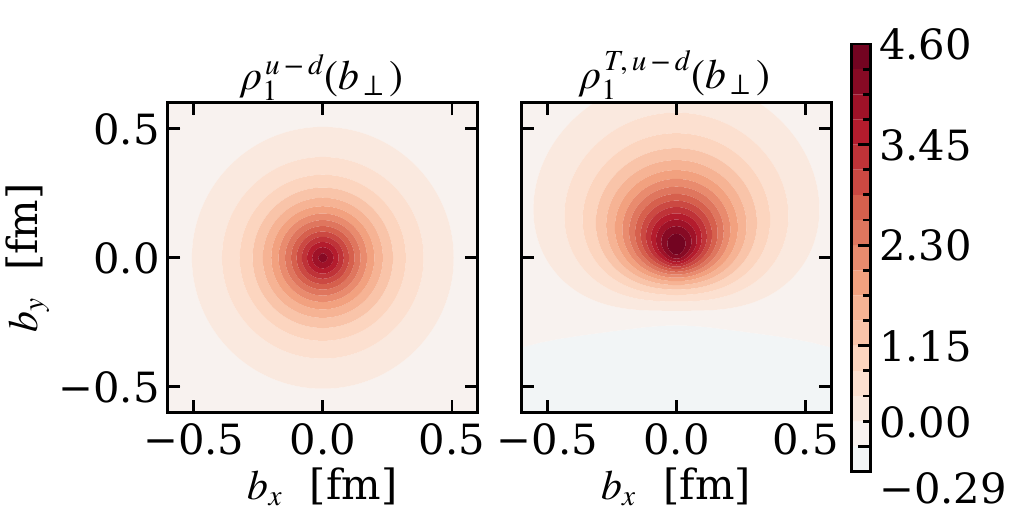}
\includegraphics[width=\columnwidth]{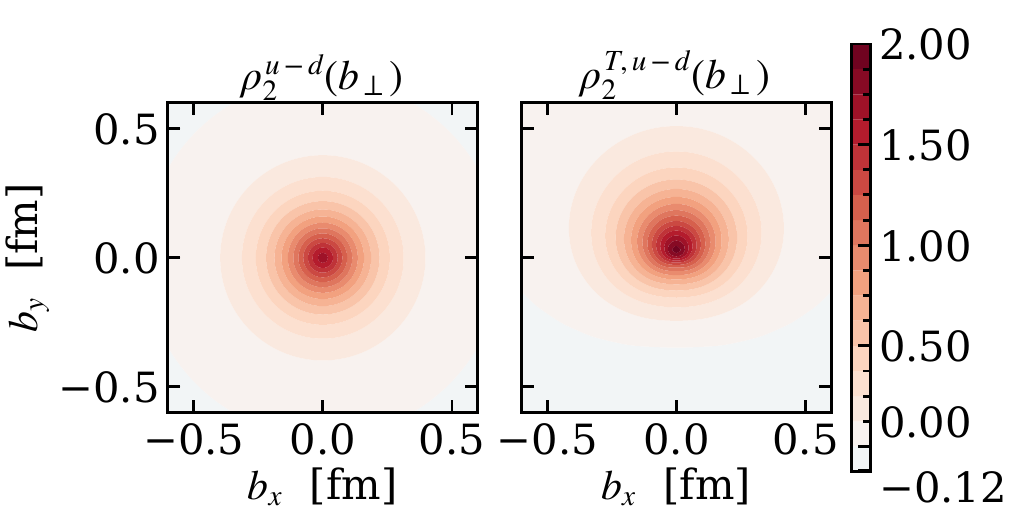}
\includegraphics[width=\columnwidth]{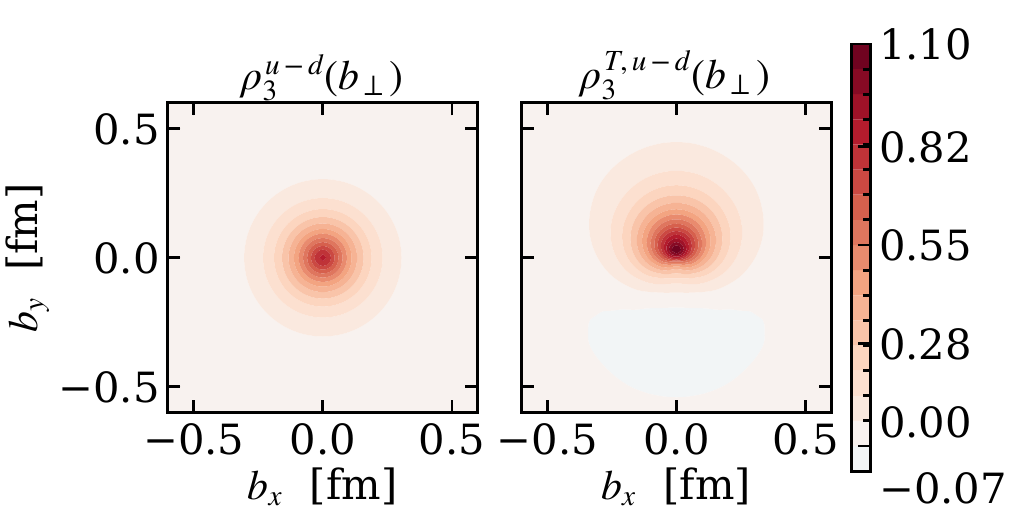}
\caption{Impact-parameter distributions of Eqs.~\eqref{eq:density-unpol} and \eqref{eq:density-pol} for $n = 0$, 1 and 2 from top to bottom, in fm$^{-2}$. The left panels show the distribution in an unpolarized proton and the right panels the one in a proton transversely polarized along $\hat{x}$. Since the $u$ and $d$ quark distributions are shifted in opposite directions in a transversely polarized proton, $\rho^{T}_{n+1}$ becomes negative at $b_y < 0$, where the $d$ quarks dominate. The GFFs are continued beyond $-t = 2.08$~GeV$^2$ with the $z$ expansion of Eq.~\eqref{eq:zexp} before the transform is taken. Each row has its own color scale.}
\label{fig:tomography}
\end{figure}

The integrations in Eqs.~\eqref{eq:density-unpol} and ~\eqref{eq:density-pol} are performed over $\vec\Delta_\perp$ from 0 to $\infty$, whereas the lattice data are available only up to $-t \le 2.08$~GeV$^2$, so the GFFs have to be continued to larger $|t|$ before the transform is taken. Since the ANNs are trained exclusively on data within $-t \le 2.08$~GeV$^2$, evaluating them beyond this boundary leads to an unphysical growth of the statistical uncertainties. We therefore continue the GFFs with a $z$ expansion~\cite{Boyd:1994tt,Hill:2010yb,Bourrely:2008za},
\begin{equation}
F(t) = \sum_{k=0}^{k_{\max}} a_k\, z^k, \qquad z(t) = \frac{\sqrt{t_{\rm cut}+|t|}-\sqrt{t_{\rm cut}}}{\sqrt{t_{\rm cut}+|t|}+\sqrt{t_{\rm cut}}},
\label{eq:zexp}
\end{equation}
with $t_{\rm cut} = 4m_\pi^2$, $m_\pi = 139.57$~MeV, and $k_{\max} = 7$, on which we impose the four conditions that $F$ and its first three derivatives with respect to $z$ vanish at $z = 1$. This enforces the expected power-law falloff at large $|t|$ and leaves four free coefficients. These are fitted to the central values of the extracted GFFs, weighted with their statistical uncertainties, for $-t \ge 0.5$~GeV$^2$ ($n = 0, 1$) and $-t \ge 0.25$~GeV$^2$ ($n = 2$). We prefer the $z$ expansion to a dipole form because it is more general. A dipole imposes a fixed functional form and a fixed sign on the continued GFF, which matters for example for $A_{3,0}$, whose central values approach zero at the edge of the data. The distributions in Fig.~\ref{fig:tomography} are computed from the central values of the GFFs, using the lattice results for $-t \le 2.08$~GeV$^2$ and the $z$ expansion beyond.

Fig.~\ref{fig:tomography} shows the resulting distributions for $n=0$, $1$ and $2$. They have the shape expected from the form factors. The unpolarized distributions are axially symmetric and become narrower with increasing $n$, since the higher moments emphasize the large-$x$ region, which is localized closer to the center. The transversely polarized distributions are shifted along $\hat{y}$ by an amount set by the size of $B_{n+1,0}$ relative to $A_{n+1,0}$, and the shift is largest for $n=0$, where $B_{1,0}(0)=\kappa_u-\kappa_d$ is largest.
Negative values appear for $b_y<0$, since the up and the down quark distributions move in opposite directions, towards positive and negative values of $b_y$, respectively.

\section{Systematic Studies}
\label{sec:systematics}

The results of Sec.~\ref{sec:results} depend on the truncation of the Ioffe-time expansion, the perturbative prescription, the network hyperparameters, and the selection of the training data. We now examine the effect of each of them in turn. It should be stressed that a precise determination of the systematic uncertainties is beyond the scope of this work, and that this section is not a rigorous model-selection procedure: the present study is a proof of concept, in which the kinematic coverage of the data limits how strongly the ANN can be constrained, as will be apparent in Sec.~\ref{sec:sys-ann}. An extension of this analysis to a broader dataset is planned for future work. Starting from the baseline analysis of Sec.~\ref{sec:results-fs}, that is $z/a=2,3$, $-t\leq2.08~{\rm GeV}^2$, $M=5$ neurons and the full data set, we change one ingredient at a time and repeat the full procedure, including the retraining of all fourteen networks on every bootstrap sample. 

The lowest order GFFs, \(A_{1,0}\) and \(B_{1,0}\), are generally so well constrained that they exhibit little sensitivity to most of the variations considered, the truncation of the Ioffe-time expansion in Sec.~\ref{sec:sys-nmax} being the main exception, and therefore provide limited discrimination between different analysis choices. We instead display the lowest order GFFs for which a systematic dependence becomes visible while the extraction remains sufficiently constrained for the comparison to be meaningful. These are \(A_{2,0}\), \(B_{2,0}\), \(A_{3,0}\) and \(B_{3,0}\), which we show in all comparisons of this section, with \(A_{2,0}\) and \(B_{2,0}\) in the upper and \(A_{3,0}\) and \(B_{3,0}\) in the lower panels, except for the truncation test of Sec.~\ref{sec:sys-nmax} and the determination of $C_2$ and $C_4$ in Sec.~\ref{sec:sys-dterm}, which concern other GFFs. The higher-order GFFs that are not shown here are less well constrained, so their larger variations cannot be cleanly attributed to the systematic under study. In all figures in this section, the error bands denote statistical uncertainties only, and shifts quoted in standard deviations refer to the statistical uncertainty of the baseline unless stated otherwise. Moreover, we draw the baseline as a red outline, and the variations as filled bands, so the baseline remains visible where the bands overlap.

We distinguish two kinds of variation. Within the FS prescription, the systematic uncertainty captures the difference between the two forms of the Wilson coefficients and the dependence on the network architecture. Within the RGR prescription, the systematic uncertainty is the average of the absolute differences between the results at $\kappa=1$ and $\kappa=2$ and the baseline $\kappa=\sqrt{2}$, added in quadrature with the statistical uncertainty to give the RGR results in Table~\ref{tab:fs-rgr} and the dashed frames in Fig.~\ref{fig:gff-fs}.

\subsection{Dependence on the network architecture}
\label{sec:sys-ann}

Since the GFFs are extracted by a neural network, we examine the ingredients associated with the architecture and with the training. The learning rate sets the size of the parameter updates performed by the optimizer and therefore controls both the speed and stability of the training. If it is too small, convergence may not be reached within the chosen number of iterations, whereas excessively large values can lead to oscillations or numerical instability. The number of neurons fixes how flexible the parametrization is, and has to be large enough not to bias the result and small enough not to follow the fluctuations of the data. The initialization of part of the parameters, namely the weights $W_1$ and $W_2$, is random, so the fit is repeated with several seeds to check that the result does not depend on the starting point. All comparisons in this subsection are carried out within the FS prescription.

\begin{figure}[t]
\centering
\includegraphics[width=0.4943\columnwidth]{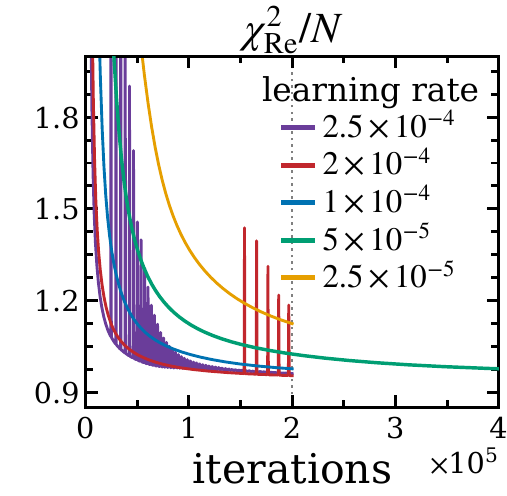}
\includegraphics[width=0.4943\columnwidth]{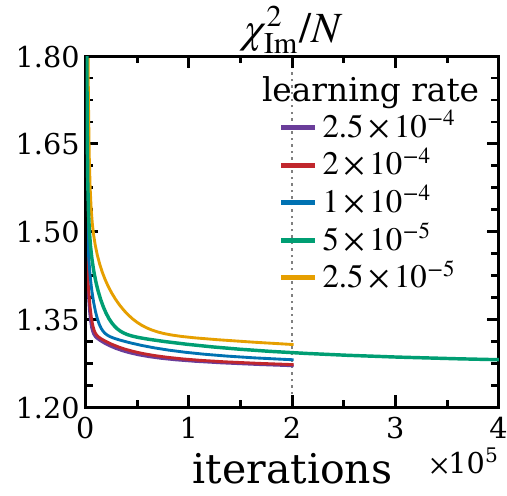}
\caption{Evolution of the two terms of the loss function of Eq.~\eqref{eq:chi2} during the training, for five learning rates, for the real (left) and the imaginary (right) part of the reduced ITD, each normalized by the number of training points $N$ entering it. The first iterations, where $\chi^{2}/N$ is of order $10^{3}$, lie above the vertical range. The training is stopped at $2\times10^{5}$ iterations, marked by the dotted line, except for the reference value $\eta=5\times10^{-5}$, which is continued to $4\times10^{5}$ iterations to check the convergence.}
\label{fig:sys-lr}
\end{figure}

The learning rate is examined first. Figure~\ref{fig:sys-lr} shows the evolution of the two contributions to the loss function in Eq.~\eqref{eq:chi2} with the number of training steps, for five learning rates and bootstrap sample 0. We focus on the real-part contribution, which exhibits the strongest dependence on the learning rate. The two largest rates do not lead to a steady descent of the loss: at $\eta=2.5\times10^{-4}$ it exhibits pronounced oscillations between about $2\times10^{4}$ and $6\times10^{4}$ iterations, and at $\eta=2\times10^{-4}$ isolated spikes appear late in the training, between about $1.5\times10^{5}$ and $1.9\times10^{5}$ iterations. This indicates that these rates are close to the stability limit of the optimization, with parameter updates that are too large to ensure a smooth descent of the loss. At $\eta=2.5\times10^{-5}$, the descent is stable but slow, and after $2\times10^{5}$ iterations the loss remains $10\%$ to $15\%$ above the values reached with the larger stable learning rates, so that the adopted training length is insufficient for this value of $\eta$. At $\eta=1\times10^{-4}$ and $\eta=5\times10^{-5}$, the loss decreases faster and still smoothly. To check the convergence, the training with $\eta=5\times10^{-5}$ has been continued to $4\times10^{5}$ iterations: over the additional $2\times10^{5}$ iterations $\chi^{2}_{\rm Re}/N$ decreases from $1.024$ to $0.976$ and $\chi^{2}_{\rm Im}/N$ from $1.293$ to $1.281$, which are approximately the values that $\eta=1\times10^{-4}$ reaches after $2\times10^{5}$ iterations, so that the two rates approach the same plateau. Over the same interval, the GFFs change by less than $0.8$ standard deviations for $A_{1,0}$, $A_{2,0}$, $B_{2,0}$, $A_{3,0}$ and $C_2$ and by about one standard deviation for $B_{1,0}$, while the largest changes, $2.2$ and $1.6$ standard deviations for $B_{3,0}$ and $C_4$, occur at the largest $|t|$, where these GFFs are least constrained. Since the loss behavior is not necessarily the same for each bootstrap replica, stability for a single sample does not by itself guarantee stable optimization for every replica. We therefore adopt $\eta=5\times10^{-5}$ as the reference value, which keeps a safety margin from the oscillatory behavior observed at the two largest rates while reaching, within the adopted training length, a loss close to its plateau. The imaginary-part contribution exhibits the same qualitative behavior and does not alter this choice.

The remaining ingredients are compared at the level of the GFFs. As in the rest of this section, we show $A_{2,0}$ and $B_{2,0}$ in the upper and $A_{3,0}$ and $B_{3,0}$ in the lower panels. The dependence on the architecture is most clearly visible for the latter, which are still determined with a useful precision. The same comparisons have been performed for all fourteen GFFs and lead to the same conclusions.

\begin{figure}[t]
\centering
\includegraphics[width=0.4943\columnwidth]{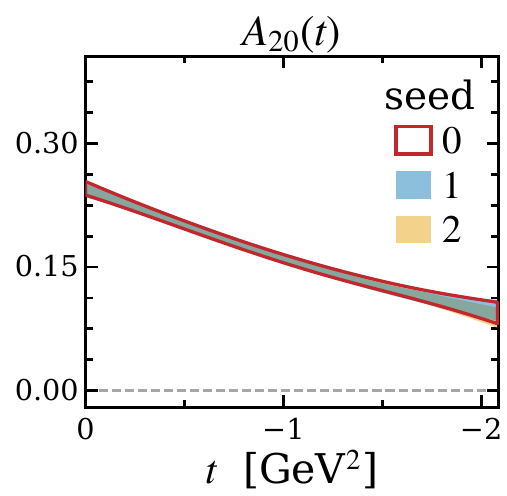}
\includegraphics[width=0.4943\columnwidth]{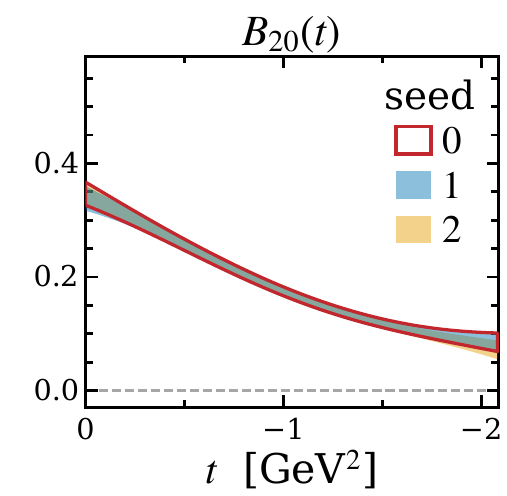}
\includegraphics[width=0.4943\columnwidth]{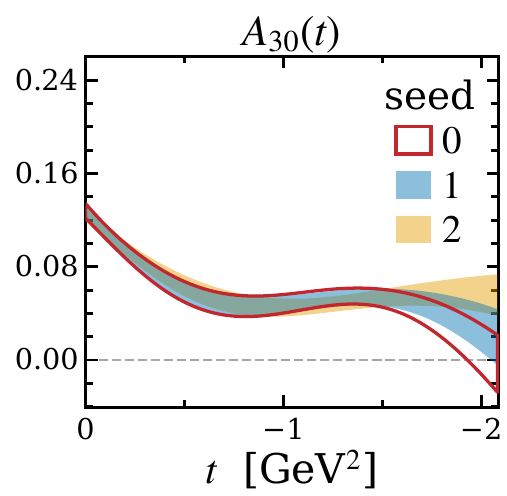}
\includegraphics[width=0.4943\columnwidth]{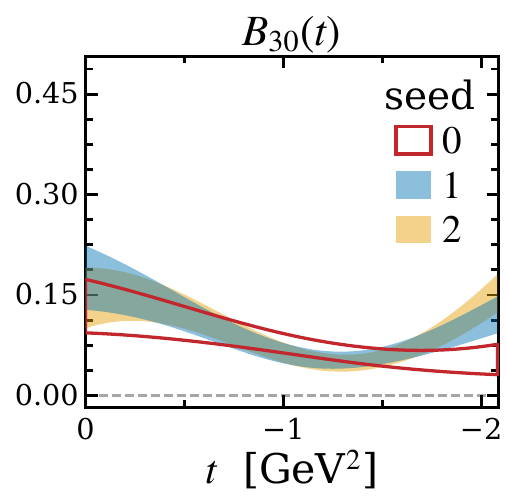}
\caption{Dependence of $A_{2,0}$, $B_{2,0}$ (upper panels) and $A_{3,0}$, $B_{3,0}$ (lower panels) on the initialization point of the network parameters. The red outline is the baseline seed and the filled bands are the two other seeds. All other analysis choices are those of the baseline. The bands denote statistical uncertainties only.}
\label{fig:sys-seed}
\end{figure}

Fig.~\ref{fig:sys-seed} shows the dependence on the initialization point. The three seeds give similar $t$ dependences, but their central values differ from the baseline by up to about $2.4$ standard deviations for $A_{3,0}$ and $4.4$ for $B_{3,0}$. For $A_{1,0}$ and $B_{1,0}$ the deviations reach $2.0$ and $4.4$ standard deviations, whereas $A_{2,0}$ and $B_{2,0}$ change by less than one standard deviation (upper panels of Fig.~\ref{fig:sys-seed}). In all cases the largest deviations occur at the largest $|t|$, where the data constrain the GFFs least, and since every bootstrap replica starts from the same initialization, the statistical uncertainty does not include initialization dependence. The skewness-dependent GFFs $A_{3,2}$, $B_{3,2}$, $A_{4,2}$ and $B_{4,2}$, $C_2$ and $C_4$, are not well constrained by the data. Consequently, their central curves show a stronger dependence on the initialization, accompanied by larger statistical uncertainties. Since these variations primarily reflect the limited constraining power of the data, they are not shown. The dependence on the initialization is therefore not negligible at the present precision, and it is included in the systematic uncertainty as described below. 

\begin{figure}[t]
\centering
\includegraphics[width=0.4943\columnwidth]{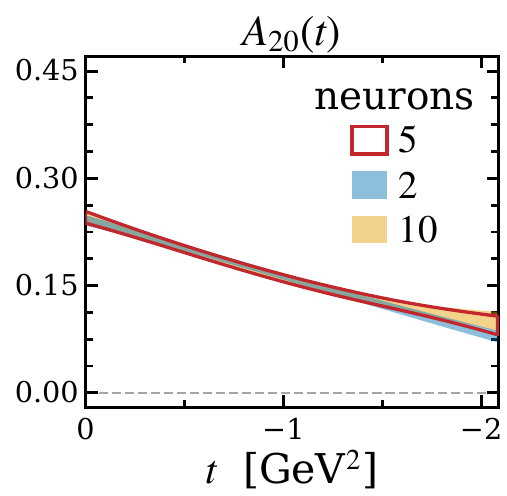}
\includegraphics[width=0.4943\columnwidth]{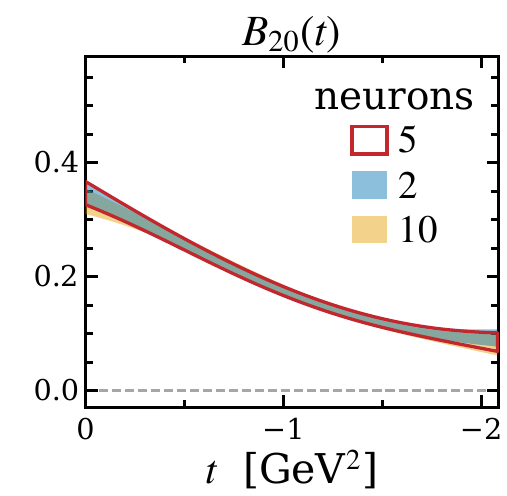}
\includegraphics[width=0.4943\columnwidth]{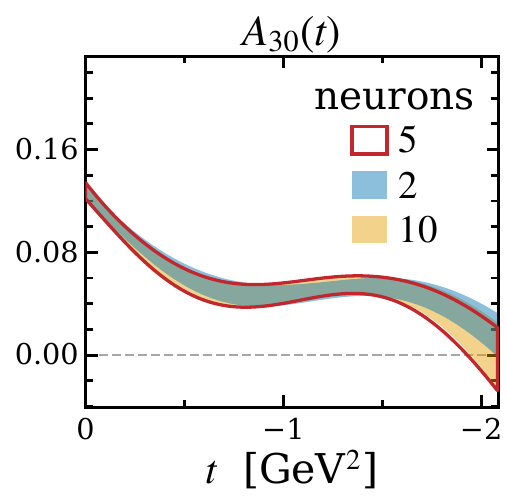}
\includegraphics[width=0.4943\columnwidth]{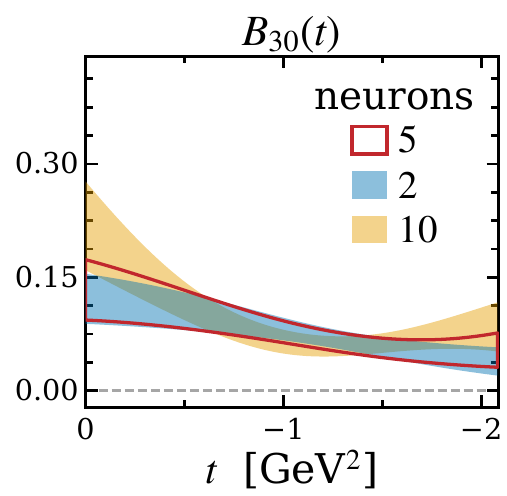}
\caption{Dependence of $A_{2,0}$, $B_{2,0}$ (upper panels) and $A_{3,0}$, $B_{3,0}$ (lower panels) on the number of neurons of the hidden layer. The red outline is the baseline $M=5$ and the filled bands are $M=2$ and $M=10$. All other analysis choices are those of the baseline.}
\label{fig:sys-neuron}
\end{figure}

Fig.~\ref{fig:sys-neuron} shows the dependence on the number of neurons. The three determinations agree within about two standard deviations, but the band widths differ, most clearly for $B_{3,0}$ at small $|t|$, where the $M=2$ band is the narrowest and the $M=10$ band the widest, with the baseline in between. The narrower band obtained with $M=2$ should not be interpreted as a more precise determination. A network with limited capacity restricts the family of functions accessible to the fit and can therefore suppress the variation among bootstrap replicas, leading to artificially narrow uncertainties even when the fit quality itself remains acceptable. This behavior is most apparent for weakly constrained GFFs such as $C_4$, for which the $M=2$ band is narrower than the variation induced by changing the initialization seed, indicating that the bootstrap uncertainty alone does not capture the full instability of the extraction in this case. Increasing the width to $M=10$ adds substantially more flexibility and broadens the uncertainty bands over part of the $t$ range, without systematically improving the data fit, as reflected in the comparable $\chi^2$ values. We therefore adopt $M=5$ as a conservative choice that provides sufficient flexibility while avoiding unnecessary model complexity.

The number of neurons visibly affects only the GFFs that the data constrain least, such as $C_4$, whereas the initialization also affects some of the better determined GFFs at large $|t|$, most notably $B_{1,0}$ and $B_{3,0}$. The present data therefore cannot single out a preferred architecture, and we do not attempt to optimize it further. This is one of the respects in which the present work is a proof of concept. With the larger and denser data sets planned for the future, a proper model-selection procedure will become both possible and necessary. We instead take the spread of the variations as an estimate of the uncertainty associated with the parametrization. For each of the two ingredients we average the absolute difference from the baseline over its two variations, that is over the two additional seeds and over $M=2$ and $M=10$, and the two averages enter the error budget as independent contributions. The learning rate does not enter this estimate, since the rates other than $1\times10^{-4}$ and $5\times10^{-5}$ are excluded by the behavior of the training itself rather than by a difference in the result.

\subsection{The fit window of Wilson line length}
\label{sec:sys-z}

The choice of the separations entering the fit is constrained from both sides. At separations approaching the lattice spacing the matrix elements receive discretization corrections that are not accounted for in Eqs.~\eqref{eq:designReH}--\eqref{eq:designImE}. At $z/a=1$ the operator is only one link long, so such effects are expected to be largest there. At long distances the relation of Eq.~\eqref{eq:matching-moments} receives power corrections of order $z^2\Lambda_{\rm QCD}^2$, $z^2m^2$ and $z^{2}t$, and in the RGR prescription the coupling at the matching scale grows. With $a=0.093$~fm, the separations $z/a=1,2,3,4$ correspond to
$z\simeq0.09$, $0.19$, $0.28$, and $0.37$~fm, respectively. Theoretical arguments alone do not tell us where the two effects become important, so we select the window by comparing fits performed with different sets of separations, as follows.

\begin{figure}[t]
\centering
\includegraphics[width=0.4943\columnwidth]{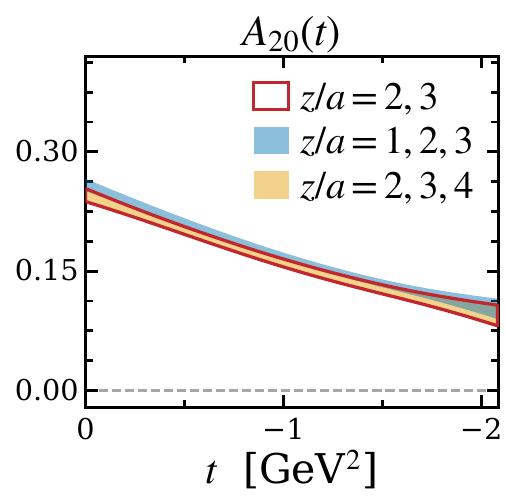}
\includegraphics[width=0.4943\columnwidth]{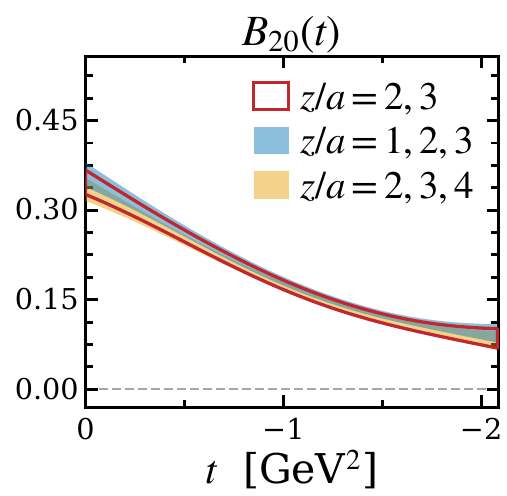}
\includegraphics[width=0.4943\columnwidth]{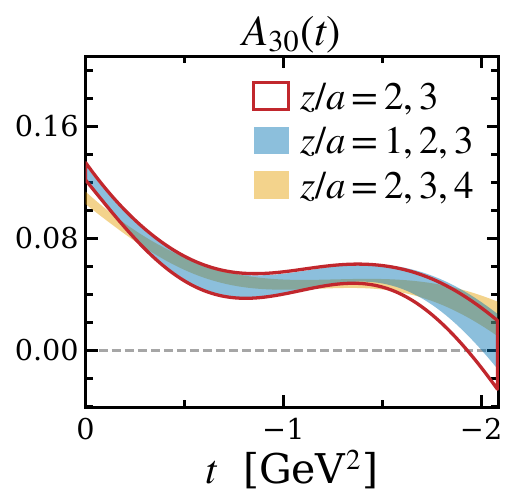}
\includegraphics[width=0.4943\columnwidth]{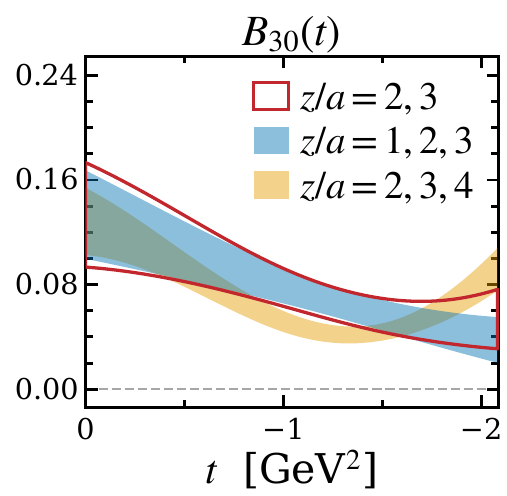}
\caption{Dependence of $A_{2,0}$, $B_{2,0}$ (upper panels) and $A_{3,0}$, $B_{3,0}$ (lower panels) on the window of Wilson line lengths entering the fit. The red outline is the baseline window $z/a=2,3$ and the filled bands are obtained by adding $z/a=1$ and $z/a=4$. All other analysis choices are those of the baseline. The bands denote statistical uncertainties only.}
\label{fig:sys-z}
\end{figure}

Fig.~\ref{fig:sys-z} compares the baseline window with the two extensions. In the FS prescription the baseline window gives $\chi^{2}_{\rm Re}/N=1.024$ and $\chi^{2}_{\rm Im}/N=1.293$. Adding $z/a=1$ makes the fit noticeably worse, increasing these values to $1.146$ and $1.539$, respectively, which is the behavior expected if the shortest separation carries corrections that Eqs.~\eqref{eq:designReH}--\eqref{eq:designImE} do not describe. Adding $z/a=4$ possibly exposes the fit to the power corrections discussed above, and the analysis shows that it changes the $\chi^{2}$ only slightly, to $1.032$ and $1.350$, so the data description does not improve even though it increases the size of the training dataset. The GFFs move accordingly: adding $z/a=1$ shifts $A_{2,0}$ by up to $2.2$ standard deviations, and adding $z/a=4$ shifts $A_{3,0}(0)$ by $3.1$ standard deviations, while all other shifts stay below two standard deviations. Therefore, we exclude both separations from the training dataset, since these $z$-window extensions are expected to result in effects that the fitting equations do not account for.

\subsection{Momentum transfer cut}
\label{sec:sys-tcut}

\begin{figure}[tp]
\centering
\includegraphics[width=0.4943\columnwidth]{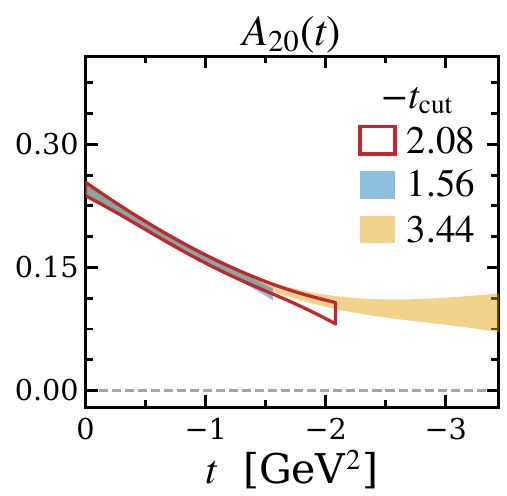}
\includegraphics[width=0.4943\columnwidth]{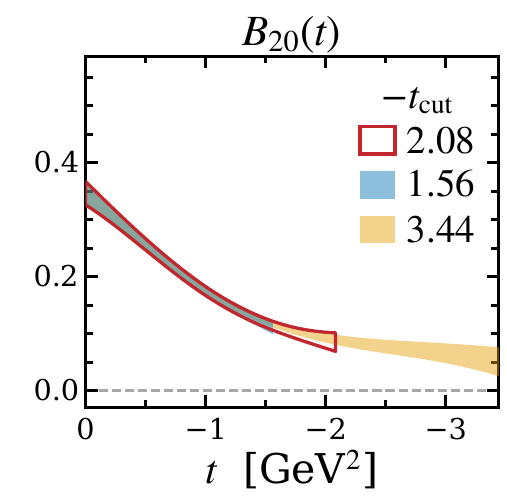}
\includegraphics[width=0.4943\columnwidth]{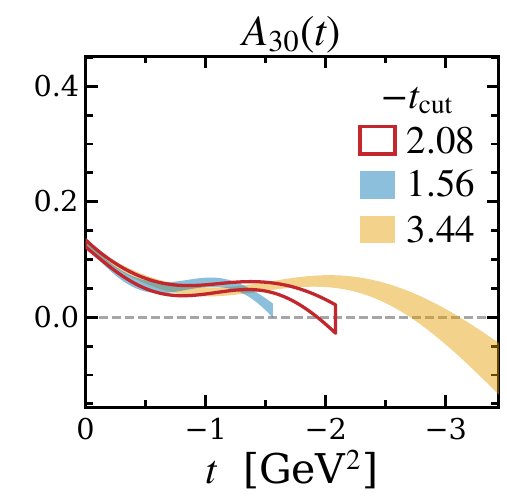}
\includegraphics[width=0.4943\columnwidth]{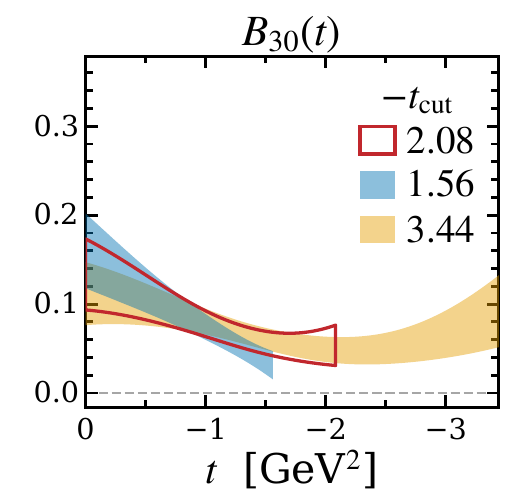}
\caption{Dependence of $A_{2,0}$, $B_{2,0}$ (upper panels) and $A_{3,0}$, $B_{3,0}$ (lower panels) on the cut in the momentum transfer. The red outline is the baseline cut $-t\leq2.08~{\rm GeV}^2$, which retains $90\%$ of the points, and the filled bands correspond to $-t\leq1.56~{\rm GeV}^2$ ($80\%$ of the points) and $-t\leq3.44~{\rm GeV}^2$ (the full data set). The horizontal range covers the full extent of the data set.}
\label{fig:sys-tcut}
\end{figure}

The short-distance relation of Eq.~\eqref{eq:matching-moments} receives kinematic power corrections of order $z^{2}t$~\cite{Braun:2026rgj}, which grow with $|t|$ and therefore limit the momentum transfers that can be included in the training. Moreover, only a few momentum transfers reach the largest $|t|$ values and carry the largest statistical errors, so they add little information while being most exposed to these corrections. The baseline cut $-t\leq2.08~{\rm GeV}^2$ removes the $10\%$ of the kinematic points with the largest $|t|$, as indicated in Fig.~\ref{fig:kinematic-coverage}, and we adopt it for the baseline analysis because it provides a compromise
between retaining most of the available kinematic information and reducing
sensitivity to the largest-$|t|$ region. Fig.~\ref{fig:sys-tcut} compares this cut with $-t\leq1.56~{\rm GeV}^2$, which removes $20\%$ of the points, and with $-t\leq3.44~{\rm GeV}^2$, which retains the entire data set. The cuts are defined as fractions of the data rather than by a particular value of $t_{\rm cut}$, because the $t$ values are not uniformly distributed, as Fig.~\ref{fig:kinematic-coverage} shows, so that only a cut on a fraction of the points corresponds to a controlled change of the training set. The purpose of the test is not to identify a preferred value of $t_{\rm cut}$, but to check how sensitive the extracted GFFs are to the inclusion of the largest-\(|t|\) data. 

At small $|t|$, where the forward limits are determined, the three bands agree within uncertainties, so that the largest-$|t|$ data do not affect that region. Analogous tests that remove $30\%$, $40\%$, and $50\%$ of the points are collected in Appendix~\ref{app:hyper} and lead to similar conclusions. Beyond $-t\simeq2~{\rm GeV}^2$ the loosest cut is supported by very few kinematic points and its uncertainty grows accordingly. We therefore retain the baseline cut. We do not assign a systematic uncertainty from this variation: the comparison shows that excluding the largest-$|t|$ data does not affect the results in the region where the data constrain the GFFs, while a quantitative estimate of the $z^{2}t$ corrections would require a dedicated study of the power corrections, which is beyond the scope of this work. In the present analysis, these corrections are reduced by restricting the fit to values above $t\simeq-2~{\rm GeV}^2$.

\subsection{RGR and the choice of the separation scale}
\label{sec:sys-kappa}

The RGR prescription of Eq.~\eqref{eq:rgr} performs the scale evolution by first evaluating the short-distance matching with $l=1+2\ln\kappa$ and the coupling $\alpha_s(\mu_0^2)$ at $\mu_0=2\kappa e^{-\gamma_E}/z$, where the constant $\kappa$ is not fixed by the formalism. Small values of $\kappa$ minimize the logarithm $l$ but push $\mu_0$ towards the nonperturbative region, while large values keep the coupling small at the price of a larger logarithm. For the separations used here the two effects are quantified in Table~\ref{tab:kappa-scales}. 

\begin{table}[t]
\caption{Separation scale $\mu_0=2\kappa e^{-\gamma_E}/z$ in GeV for the two Wilson line lengths of the baseline analysis, the associated logarithm $l=1+2\ln\kappa$, and the corresponding values of the strong coupling, for the five values of $\kappa$ considered. The coupling is evolved to $\alpha_s(2\,{\rm GeV})=0.393945$~\cite{FlavourLatticeAveragingGroupFLAG:2024oxs}. For comparison, in the fixed-scale analysis $L=\ln\!\left(e^{2\gamma_E+1}z^2\mu^2/4\right)=2.04$ and $2.85$ for $z/a=2$ and $3$, respectively.}
\label{tab:kappa-scales}
\begin{ruledtabular}
\begin{tabular}{cccccc}
 & & \multicolumn{2}{c}{$z/a=2$} & \multicolumn{2}{c}{$z/a=3$} \\
\cline{3-4} \cline{5-6}
$\kappa$ & $l$ & $\mu_0$ & $\alpha_s(\mu_0)$ & $\mu_0$ & $\alpha_s(\mu_0)$ \\
\colrule
$1/\sqrt{2}$ & $0.31$ & $0.84$ & $0.72$ & $0.56$ & $1.17$ \\
$1$          & $1.00$ & $1.19$ & $0.54$ & $0.79$ & $0.76$ \\
$\sqrt{2}$   & $1.69$ & $1.68$ & $0.43$ & $1.12$ & $0.56$ \\
$2$          & $2.39$ & $2.38$ & $0.36$ & $1.59$ & $0.45$ \\
$2\sqrt{2}$  & $3.08$ & $3.37$ & $0.31$ & $2.25$ & $0.37$
\end{tabular}
\end{ruledtabular}
\end{table}

\begin{figure}[t]
\centering
\includegraphics[width=0.4943\columnwidth]{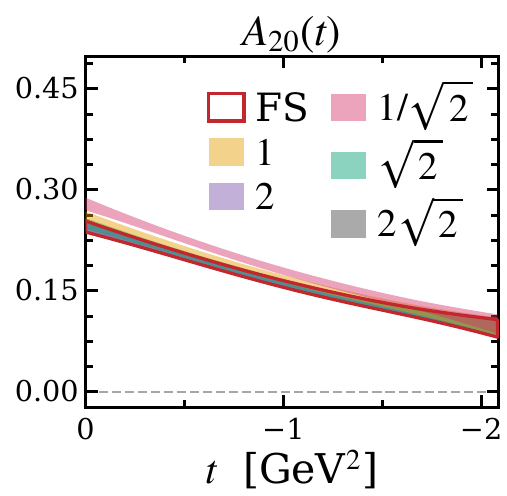}
\includegraphics[width=0.4943\columnwidth]{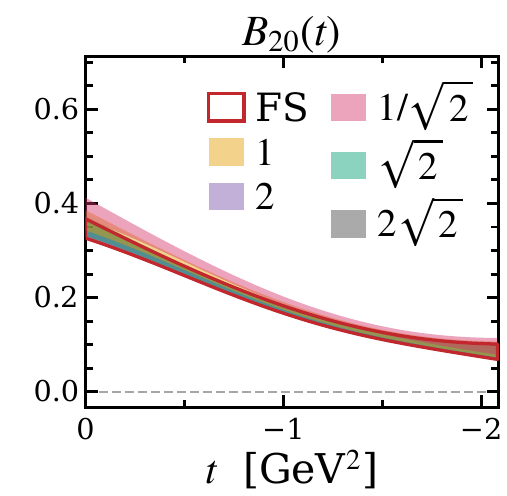}
\includegraphics[width=0.4943\columnwidth]{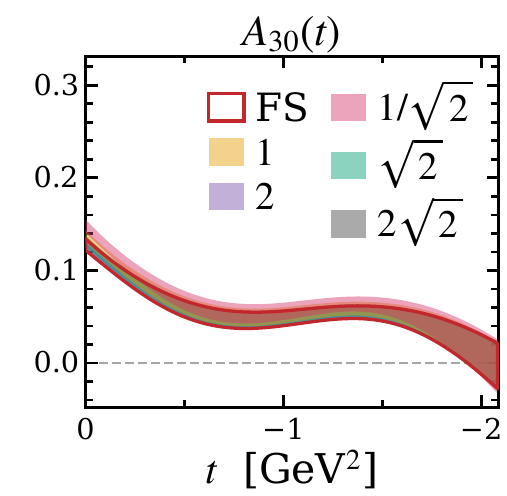}
\includegraphics[width=0.4943\columnwidth]{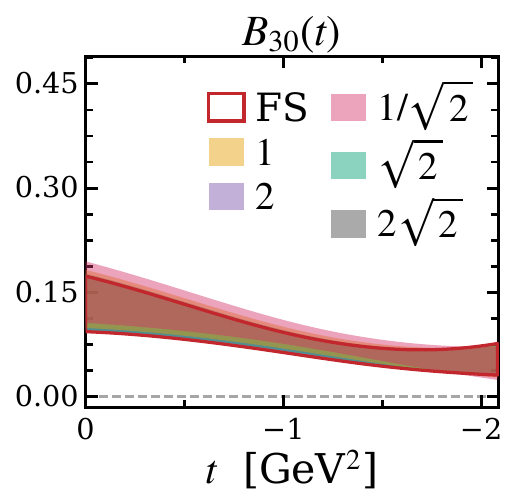}
\caption{Dependence of $A_{2,0}$, $B_{2,0}$ (upper panels) and $A_{3,0}$, $B_{3,0}$ (lower panels) on the choice of $\kappa$. The red outline is the fixed energy scale baseline and the filled bands are the RGR results for $\kappa=1/\sqrt{2}$, $1$, $\sqrt{2}$, $2$ and $2\sqrt{2}$. The corresponding matching scales are listed in Table~\ref{tab:kappa-scales}.}
\label{fig:sys-kappa}
\end{figure}

\begin{figure}[t]
\centering
\includegraphics[width=\columnwidth]{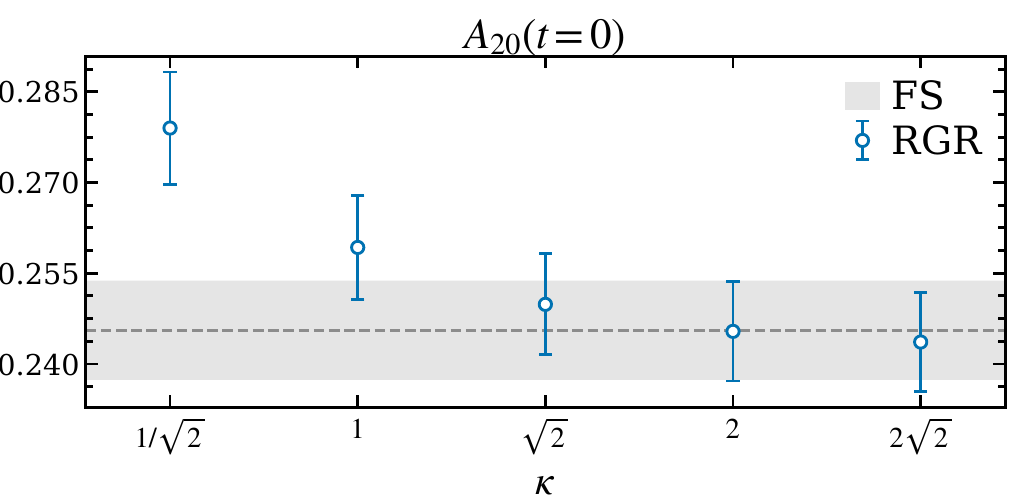}
\includegraphics[width=\columnwidth]{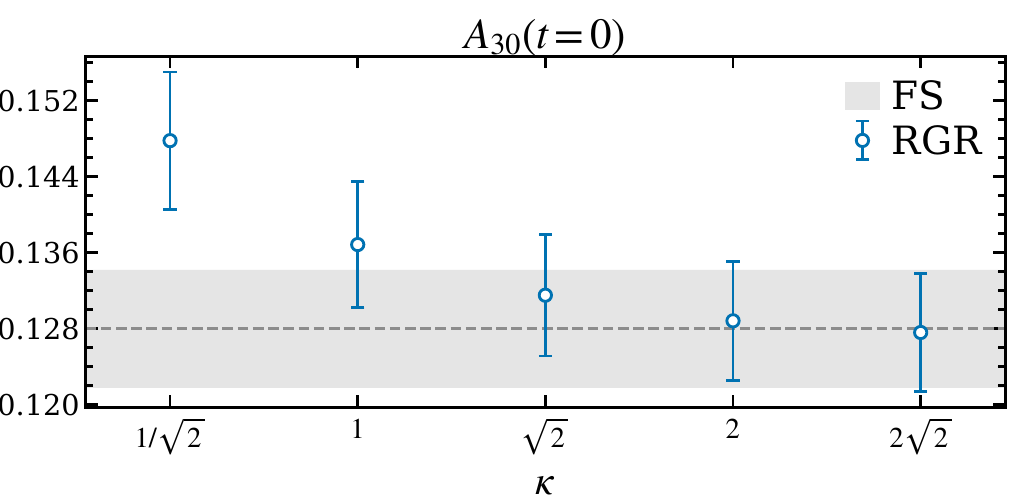}
\caption{Forward values $A_{2,0}(t=0)$ (upper panel) and $A_{3,0}(t=0)$ (lower panel) as a function of the constant $\kappa$ that fixes the matching scale $\mu_0=2\kappa e^{-\gamma_E}/z$. The points are the RGR determinations with their statistical uncertainties and the horizontal band is the fixed energy scale (FS) result. The horizontal axis is logarithmic in $\kappa$.}
\label{fig:sys-kappa-a20}
\end{figure}

Fig.~\ref{fig:sys-kappa} shows $A_{2,0}$, $B_{2,0}$, $A_{3,0}$ and $B_{3,0}$ for the five values of $\kappa$ together with the fixed energy scale result. The dependence on $\kappa$ is monotonic, and the smallest values of $\kappa$, where the matching scale of Table~\ref{tab:kappa-scales} is lowest, show the largest deviation from the FS determination. This is more explicitly shown in Fig.~\ref{fig:sys-kappa-a20}, which shows the forward values $A_{2,0}(t=0)$ and $A_{3,0}(t=0)$ as a function of $\kappa$. For both, the RGR results decrease as $\kappa$ is increased from $1/\sqrt{2}$ and,
for $\kappa\gtrsim\sqrt{2}$, become consistent with the FS result within
uncertainties. The behavior at the two smallest values is consistent with
a loss of perturbative control as the matching scale is lowered. At
$\kappa=1/\sqrt{2}$, the longer separation is matched at
$\mu_0\simeq0.56$~GeV, where the coupling becomes large and the
matching is no longer expected to be reliable. The corresponding result also
differs substantially from the FS determination. The perturbative instability
associated with $\kappa=1/\sqrt{2}$ is further discussed in Sec.~\ref{sec:sys-alpha}. At $\kappa=1$, the longer separation is matched at
$\mu_0\simeq0.79$~GeV, still close to the nonperturbative region.

We therefore adopt $\kappa=\sqrt{2}$ as the central value of the scale parameter in the RGR analysis. It lies midway in $\ln\kappa$ between $\kappa=1$ and $\kappa=2$ and avoids the very low matching scales encountered for $\kappa=1/\sqrt{2}$. Since $\kappa$ is not fixed by the formalism, the residual dependence on it is a consequence of the truncation of the perturbative series, and we convert it into a systematic uncertainty of the RGR results, taking the average of the absolute differences of $\kappa=1$ and $\kappa=2$ from $\kappa=\sqrt{2}$, as described in Sec.~\ref{sec:results-rgr}. The values $\kappa=1/\sqrt{2}$ and $2\sqrt{2}$ are not included, the former because of the loss of perturbative control discussed above.

\subsection{Form of the Wilson coefficients}
\label{sec:sys-alpha}

\begin{figure}[t]
\centering
\includegraphics[width=0.4943\columnwidth]{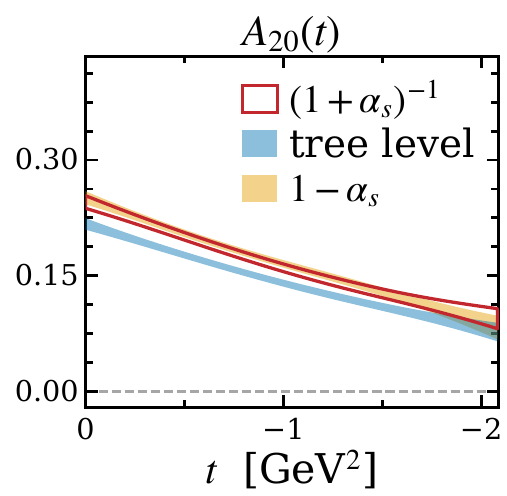}
\includegraphics[width=0.4943\columnwidth]{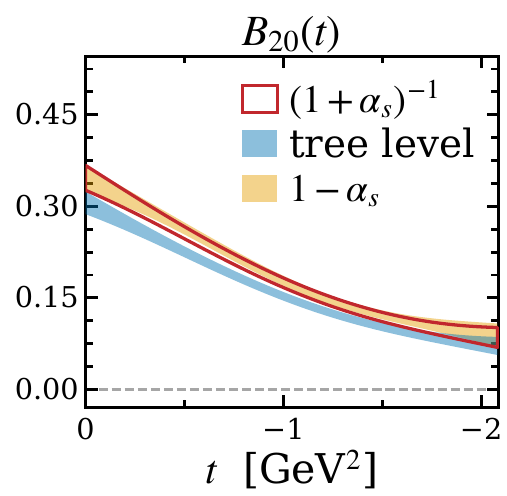}
\includegraphics[width=0.4943\columnwidth]{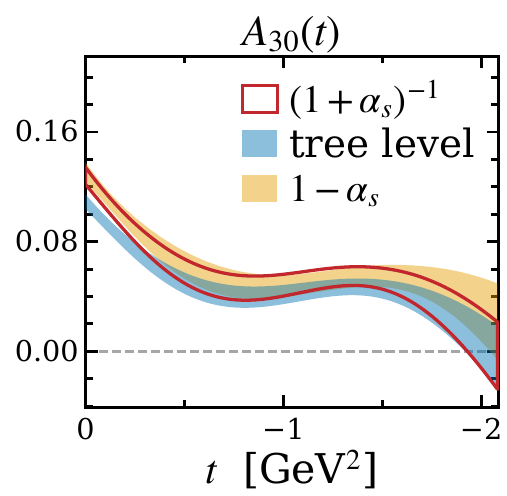}
\includegraphics[width=0.4943\columnwidth]{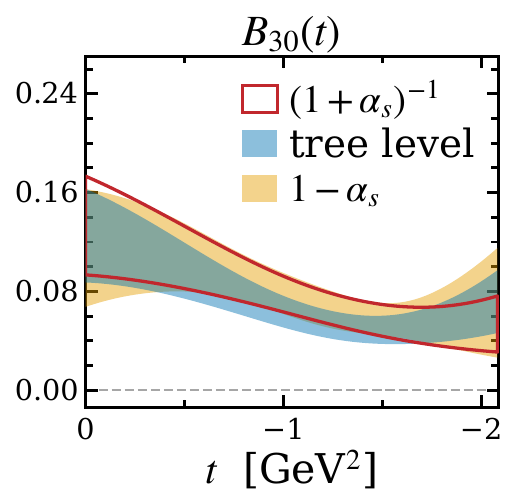}
\caption{Dependence of $A_{2,0}$, $B_{2,0}$ (upper panels) and $A_{3,0}$, $B_{3,0}$ (lower panels) on the form of the Wilson coefficients in the fixed energy scale analysis. The red outline is the baseline inverted form of Eq.~\eqref{eq:dn-fs}, and the filled bands are the tree-level choice and the expanded form $1-a_s(\gamma_{n+1}L+b_{n+1})$.}
\label{fig:sys-alpha}
\end{figure}

The relations fitted to the data involve the inverted form $c_{n+1,0}=1/[1+a_s(\gamma_{n+1}L+b_{n+1})]$ of Eq.~\eqref{eq:dn-fs}. This differs from the expanded form $1-a_s(\gamma_{n+1}L+b_{n+1})$ only at $\mathcal{O}(\alpha_s^2)$, so the two expressions are equivalent to the order at which the
matching coefficients are known. Fig.~\ref{fig:sys-alpha} compares the two, together with the tree-level choice, in which the $\mathcal{O}(\alpha_{s})$ terms of Table~\ref{tab:wilson} are switched off altogether, that is $c_{n+1,0}=1$ and $c_{n+1,2}=0$. The inverted and the expanded forms agree within about one standard deviation over the whole range of $t$, as expected from a difference of order $\alpha_s^2$. The tree-level band lies below both NLO determinations for $A_{2,0}$ and $B_{2,0}$, at $t=0$ by $3.5$ and $2.0$ standard deviations, respectively, and deviates by $3.0$ standard deviations also for $A_{3,0}$, while for $B_{3,0}$ the shift remains well within its larger statistical uncertainty, illustrating the size of the one-loop correction. The difference
between both forms of the Wilson coefficients, which first appears at
$\mathcal{O}(\alpha_s^2)$, is used as an estimate of the neglected higher-order
contributions, and its absolute value enters the systematic uncertainty of the FS analysis.
The tree-level result is shown only to illustrate the size of the NLO correction
and is not included in the error budget. A genuine assessment of the
order-by-order perturbative convergence would require an NNLO calculation.

The same comparison between the inverted and the expanded forms can be performed within the RGR prescription, using Eq.~\eqref{eq:dn}. For $\kappa\geq1$ the two forms again lead to consistent results, whereas at $\kappa=1/\sqrt{2}$ the extracted GFFs show a strong sensitivity to this choice. This is a further indication that the perturbative treatment is no longer under control at such low matching scales, and supports the choice of a larger value of $\kappa$ as the central one in Sec.~\ref{sec:sys-kappa}.

\subsection{Truncation of the Ioffe-time expansion}
\label{sec:sys-nmax}

\begin{figure}[t]
\centering
\includegraphics[width=0.4943\columnwidth]{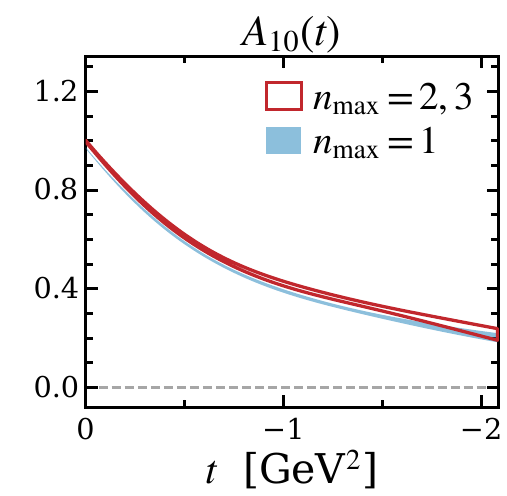}
\includegraphics[width=0.4943\columnwidth]{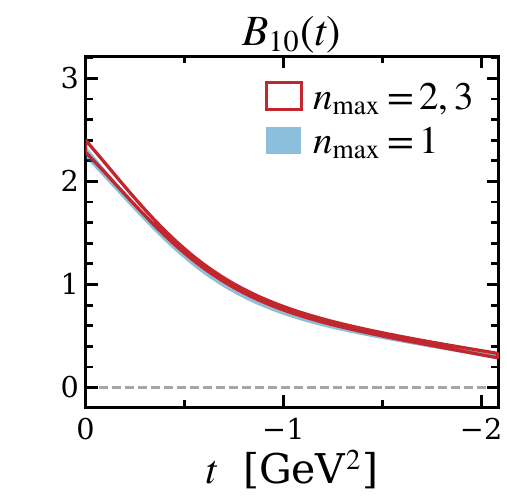}
\includegraphics[width=0.4943\columnwidth]{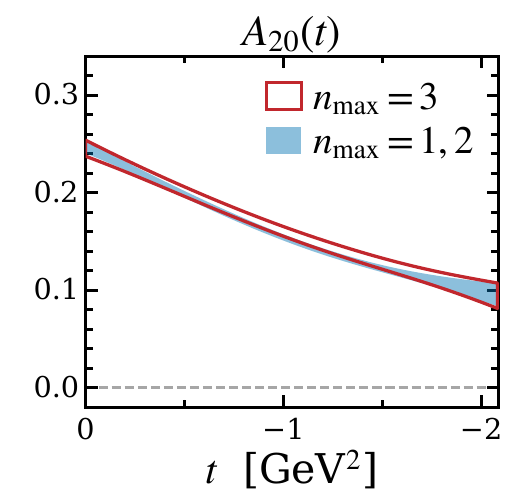}
\includegraphics[width=0.4943\columnwidth]{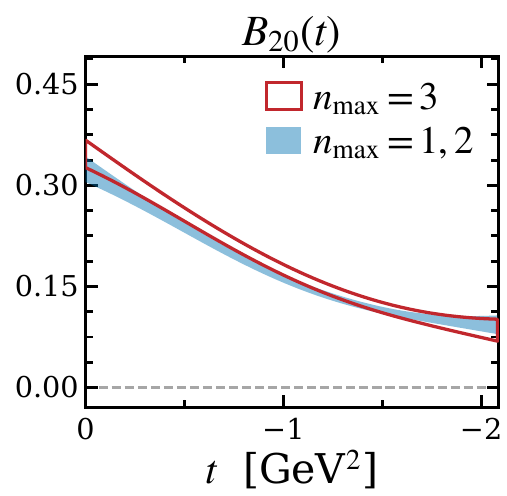}
\caption{Dependence of $A_{1,0}$, $B_{1,0}$ (upper two panels) and $A_{2,0}$, $B_{2,0}$ (lower two panels) on the order at which the Ioffe-time expansion is truncated. The red outline is the baseline $n_{\max}=3$ and the filled band the truncation at a lower order. As follows from Eqs.~\eqref{eq:nmax1}--\eqref{eq:nmax3}, the first moments are identical at $n_{\max}=2$ and $n_{\max}=3$, so that the upper panels compare $n_{\max}=2,3$ with $n_{\max}=1$, while the second moments are identical at $n_{\max}=1$ and $n_{\max}=2$, so that the lower panels compare $n_{\max}=3$ with $n_{\max}=1,2$.}
\label{fig:sys-nmax}
\end{figure}

The expansion of Eq.~\eqref{eq:nu-expansion} is truncated at $n_{\max}=3$ in the baseline analysis. Lowering the truncation removes the highest moments from the fit, so that their contribution can be partially absorbed by the lower ones. Which GFFs are affected follows directly from Eq.~\eqref{eq:nu-expansion} and Eqs.~\eqref{eq:polyH}--\eqref{eq:polyE}. At $n_{\max}=1$, the reduced ITD contains only the first two moments,
\begin{align}
\operatorname{Re}F_{n_{\max}=1} &= \overline F_1, & \operatorname{Im}F_{n_{\max}=1} &= -\nu c_{2,0}\,\overline F_2 .
\label{eq:nmax1}
\end{align}
Increasing the truncation to $n_{\max}=2$ adds the third moment,
\begin{align}
\operatorname{Re}F_{n_{\max}=2} &= \left(1-\frac{\nu^2}{2}\xi^2c_{3,2}\right)\overline F_1-\frac{\nu^2}{2}c_{3,0}\,\overline F_3, \nn\\
\operatorname{Im}F_{n_{\max}=2} &= -\nu c_{2,0}\,\overline F_2 .
\label{eq:nmax2}
\end{align}
Thus, going from $n_{\max}=1$ to $n_{\max}=2$ modifies only the real part and can therefore affect $A_{1,0}$ and $B_{1,0}$, while the extraction of $A_{2,0}$ and $B_{2,0}$ is unchanged. Increasing the truncation once more to $n_{\max}=3$ adds the fourth moment and modifies only the imaginary part,
\begin{align}
\operatorname{Im}F_{n_{\max}=3} &= -\nu\left(c_{2,0}-\frac{\nu^2}{6}\xi^2c_{4,2}\right)\overline F_2+\frac{\nu^3}{6}c_{4,0}\,\overline F_4 ,
\label{eq:nmax3}
\end{align}
while the real part remains the same as in Eq.~\eqref{eq:nmax2}. Consequently, the $n_{\max}=2\to3$ step can only affect $A_{2,0}$ and $B_{2,0}$. Fig.~\ref{fig:sys-nmax} compares these two pairs.

Each pair of truncations that is identical by construction is therefore shown once in Fig.~\ref{fig:sys-nmax}: the first-order GFFs are compared between $n_{\max}=2,3$ and $n_{\max}=1$, and the second-order GFFs between $n_{\max}=3$ and $n_{\max}=1,2$. The step from $n_{\max}=1$ to $n_{\max}=2$, which adds the $\nu^2$ term to the real part, has a significant effect on $A_{1,0}$, whose forward value moves from $0.971$ to $1.000$, that is by about ten statistical standard deviations and onto the normalization fixed by the isovector charge, while $B_{1,0}$ changes by at most $3.0$ standard deviations. The step from $n_{\max}=2$ to $n_{\max}=3$, which adds the $\nu^3$ term to the imaginary part, shifts $A_{2,0}$ and $B_{2,0}$ by at most $1.3$ and $2.0$ standard deviations, respectively, so that the corresponding bands remain compatible within about two standard deviations over the full range of $t$. As Eq.~\eqref{eq:bound} shows, the moments are bounded, so the dropped terms are controlled by the factor $\nu^n/n!$, and the small effect of the last step is consistent with this suppression over the range of Ioffe times entering the fit. The comparison does not by itself establish that $n_{\max}=3$ is sufficient, which would require going to $n_{\max}=4$ and $5$, but it shows that the truncation at $n_{\max}=1$ is inadequate for the first moments, whereas the effect of the last step is at the level of the statistical uncertainty.

\subsection{Role of the nonzero-skewness data}
\label{sec:sys-xi}

\begin{figure}[t]
\centering
\includegraphics[width=0.4943\columnwidth]{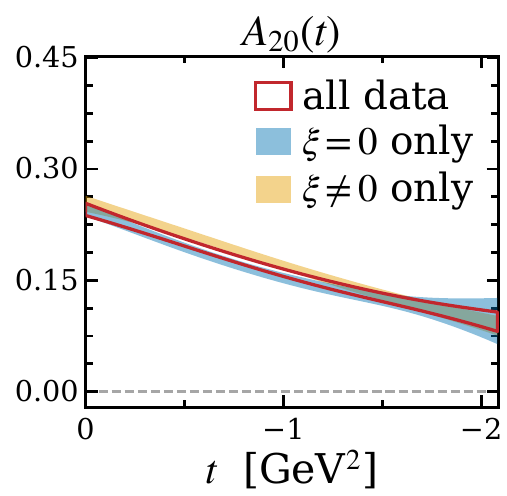}
\includegraphics[width=0.4943\columnwidth]{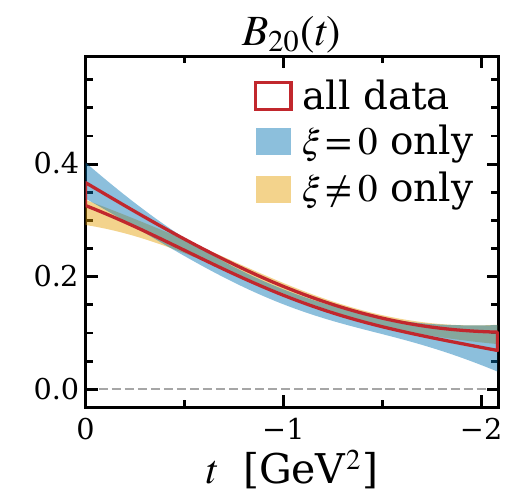}
\includegraphics[width=0.4943\columnwidth]{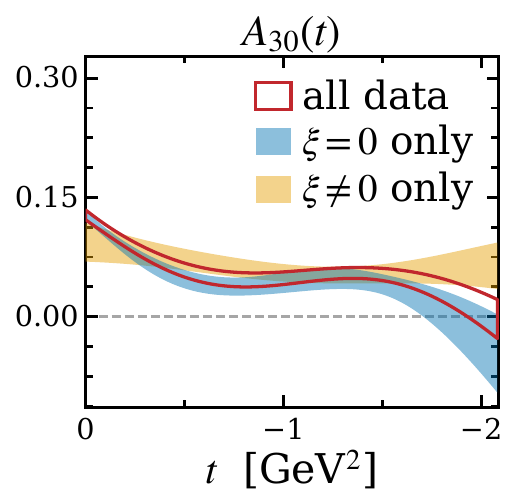}
\includegraphics[width=0.4943\columnwidth]{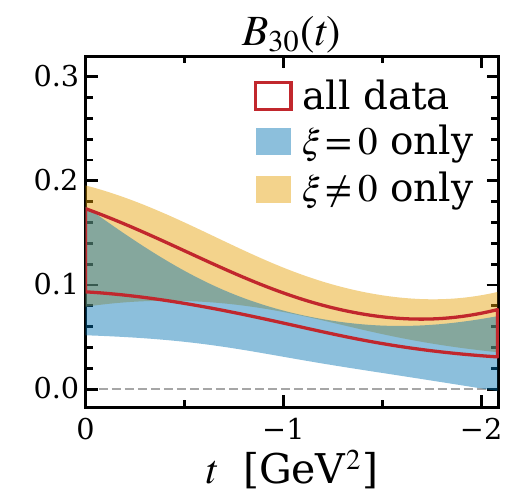}
\caption{Dependence of $A_{2,0}$, $B_{2,0}$ (upper panels) and $A_{3,0}$, $B_{3,0}$ (lower panels) on the composition of the training data. The red outline is the baseline fit to the complete set and the filled bands are the fits restricted to the zero-skewness and to the nonzero-skewness reduced ITDs.}
\label{fig:sys-xi}
\end{figure}

The zero- and nonzero-skewness data enter the same fit but constrain different combinations of GFFs. As discussed in Sec.~\ref{sec:sde}, the GFFs with $i\neq0$ require $\xi\neq0$, while those with $i=0$ are accessible already at $\xi=0$. As a consistency check, Fig.~\ref{fig:sys-xi} shows the baseline fit together with the two fits restricted to each subset. All three determinations agree within two standard deviations of their combined statistical uncertainties for all four GFFs, which shows that the two subsets are mutually consistent where both apply. Beyond this consistency check, the comparison shows what the new kinematic regime brings. Adding the nonzero-skewness data to the zero-skewness ones reduces the statistical uncertainty at $t=0$ by about $10\%$ for $A_{2,0}$ and $A_{3,0}$ and by about $35\%$ for $B_{2,0}$ and $B_{3,0}$, and on average over the fitted range of $t$ by $35\%$ to $45\%$ for all four GFFs. The gain at $t=0$ is smallest for the $A$-type GFFs, whose forward values are fixed mainly by the forward matrix elements, which belong to the zero-skewness subset, whereas no forward information is available for the $B$-type GFFs. Conversely, the nonzero-skewness data alone determine $A_{2,0}$, $B_{2,0}$ and $B_{3,0}$ at $t=0$ with uncertainties only about $10\%$ to $50\%$ larger than those of the full data set, while for $A_{3,0}$ the uncertainty grows by a factor of about four without the zero-skewness data. The nonzero-skewness data are moreover the only subset with direct sensitivity to the skewness-dependent GFFs and to \(C_2\) and \(C_4\). 

\subsection{Determination of $C_2$ and $C_4$}
\label{sec:sys-dterm}

\begin{figure}[tp]
\centering
\includegraphics[width=0.4943\columnwidth]{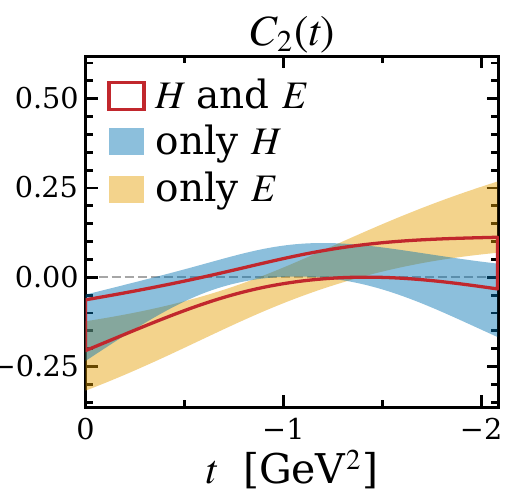}
\includegraphics[width=0.4943\columnwidth]{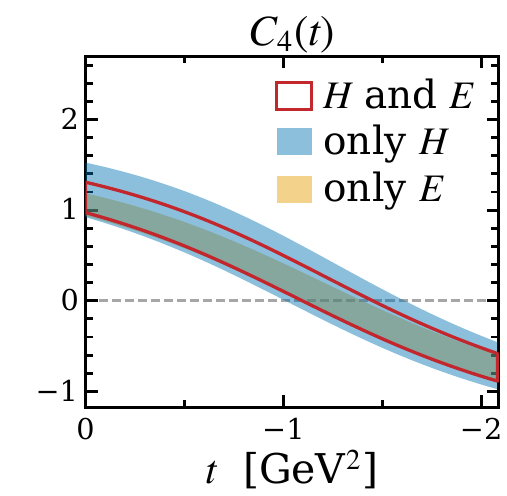}
\caption{Determination of $C_2$ (left) and $C_4$ (right). The red outline is the baseline simultaneous fit to the $H$- and $E$-type channels and the filled bands are obtained by constraining $C_2$ and $C_4$ from one channel at a time.}
\label{fig:sys-dterm}
\end{figure}

The GFFs $C_2$ and $C_4$ enter the Im $H$ and Im $E$ channels. In particular, $C_2$ appears through the combinations
$A_{2,0}+\xi^2 C_2$ and $B_{2,0}-\xi^2 C_2$, while $C_4$ enters analogously at $n=3$ through
$A_{4,0}+\xi^2 A_{4,2}+\xi^4 C_4$ and
$B_{4,0}+\xi^2 B_{4,2}-\xi^4 C_4$. In the baseline analysis $C_2$ and $C_4$ are determined from a simultaneous fit to the $H$- and $E$-type data, which uses the largest amount of information available on them. To test this choice, we repeat the fit determining $C_2$ and $C_4$ from one channel at a time, that is using only the imaginary part of $F^{H}$, where they appear together with the $A$-type GFFs, or only the imaginary part of $F^{E}$, where they appear together with the $B$-type ones. Fig.~\ref{fig:sys-dterm} shows the outcome. Their uncertainties at $t=0$ remain within a factor of two of those of the baseline, and the agreement with the baseline is noticeably worse for $C_2$ than for $C_4$. This ordering is expected: compared to \(C_2\), \(C_4\) contributes much less to both channels because of its \(\xi^4\) dependence and the corresponding higher-moment Ioffe-time prefactor. Consequently, \(C_4\) is weakly constrained by the data, and neither the \(H\) nor the \(E\) data provide substantial independent leverage on \(C_4\).

\subsection{Summary of the uncertainties}
\label{sec:sys-summary}

The variations examined above separate into two groups. Within the fixed energy scale prescription the systematic uncertainty of each GFF collects three contributions. The first is the form of the Wilson coefficients, taken as the absolute difference between the inverted and the expanded forms. The remaining two come from the network, as discussed in Sec.~\ref{sec:sys-ann}, namely from the number of neurons and from the initialization point, each taken as the average of the absolute differences between the baseline and its two variations. The three are added in quadrature with the statistical uncertainty to give the total uncertainty shown as the outer band in Fig.~\ref{fig:gff-fs}. Within the RGR prescription, the systematic uncertainty is the average of the absolute differences between the results at $\kappa=1$ and $\kappa=2$ and the baseline $\kappa=\sqrt{2}$, added in quadrature with the statistical uncertainty to give the RGR results in Table~\ref{tab:fs-rgr} and the dashed frames in Fig.~\ref{fig:gff-fs}.

\section{Conclusions}
\label{sec:conclusions}

We have presented a lattice QCD determination of proton isovector GPDs to extract the first four Mellin moments, including the skewness-dependent GFFs that are accessible only at $\xi\neq0$. The calculation combines the polynomiality of the Mellin moments in $\xi$ with short-distance factorization, therefore relating the GFFs to the reduced Ioffe-time distributions, computed in asymmetric frames of reference on an $N_f=2+1+1$ twisted-mass ensemble at $m_\pi\simeq260$~MeV. Compared with previous work, the data set is extended by a fourth sink momentum, which reaches $P_3\simeq2.08$~GeV and covers the skewness range $-1\leq\xi\leq1/2$ together with $110$ values of the momentum transfer.

The central methodological ingredient is the parametrization of each of the fourteen GFFs by an independent single-hidden-layer neural network, which takes each value of $t$ of every kinematic point as input. In asymmetric-frame kinematics, the same transverse momentum transfer corresponds to different values of $t$ at different boosts and skewness, and the network parametrization allows all these kinematics to be included in a single global fit without binning and without assuming an analytical form for the $t$ dependence. The reconstructed reduced ITDs describe the overall behavior of the lattice data across the fitted kinematic range.

The resulting GFFs display a clear hierarchy. The skewness-independent GFFs $A_{n+1,0}$ and $B_{n+1,0}$ are well determined, with the forward limits of the $A$-type GFFs constrained by the forward matrix elements. The skewness-dependent GFFs $A_{3,2}$, $B_{3,2}$, $A_{4,2}$ and $B_{4,2}$ are consistent with zero at the present precision, and $C_2$ and $C_4$ remain weakly constrained. This hierarchy follows from the suppression of these GFFs by powers of $\nu$ and $\xi$ in the fitting equations, together with the kinematic coverage of the present data. From the forward values of $A_{2,0}$ and $B_{2,0}$, we obtain $J^{u-d}=0.296(11)(7)$ at $\mu=2$~GeV in the $\overline{\rm MS}$ scheme, in agreement with previous lattice determinations based on nonlocal operators and with a total uncertainty smaller by about a factor of two than that of the corresponding zero-skewness analysis. We have also determined the impact-parameter distributions of the first three Mellin moments in an unpolarized and in a transversely polarized proton.

We have examined the dependence of the results on the network architecture and training, on the window of Wilson line lengths, on the cut in the momentum transfer, on the truncation of the Ioffe-time expansion, on the form of the Wilson coefficients, and on the scale prescription, comparing the fixed-scale analysis with a renormalization-group resummed one -- the results from the two prescriptions agree throughout the fitted $t$ interval. Since the RGR prescription introduces an auxiliary scale parameter $\kappa$ that the formalism does not fix and leaves a residual dependence at finite perturbative order, we take the fixed-scale analysis at $\mu=2$~GeV as our central result and use the RGR analysis as a complementary test of the perturbative matching.

The present calculation should be regarded as a proof of concept. Its main limitations are the use of a single ensemble, with one lattice spacing, a pion mass of about $260$~MeV and a single source-sink separation, so that neither the continuum and physical-pion-mass limits nor excited-state effects are addressed, the sparse sampling of the small-$|t|$ region, the small number of separations in the short-distance window, and the power corrections of $\mathcal{O}(z^2t)$, which have not been quantified here. Ensembles closer to the physical pion mass with several source-sink separations, larger physical volumes, finer lattice spacings, and a denser coverage in $t$ and $\xi$, together with higher-order matching and a dedicated study of the power corrections, will allow a precision determination of the skewness-dependent GFFs, including the $D$-term-related form factor $C_2$, and a systematic model selection of the network parametrization.

\begin{acknowledgments}
M.-H.~Chu, M.~Cola\c{c}o, K.~Cichy and A.~Sarkar are supported by the National Science Centre (Poland) grant OPUS No.\ 2021/43/B/ST2/00497.
M.~Constantinou acknowledges financial support by the U.S. Department of Energy, Office of Nuclear Physics, under Grant No.\ DE-SC0025218.

This research was supported in part by the PLGrid Infrastructure (Prometheus and Helios supercomputers at AGH Cyfronet in Cracow).
Computations were also partially performed at the Poznan Supercomputing and Networking Center (Eagle/Altair supercomputer), the Interdisciplinary Centre for Mathematical and Computational Modelling of the Warsaw University (Okeanos supercomputer), the Academic Computer Centre in Gda\'nsk (Tryton and Tryton Plus supercomputers) and facilities of the USQCD Collaboration, funded by the Office of Science of the U.S. Department of Energy.
This research used resources of the National Energy Research Scientific Computing Center, a DOE Office of Science User Facility supported by the Office of Science of the U.S. Department of Energy under Contract No.\ DE-AC02-05CH11231 using NERSC award NP-ERCAP0022961.
The gauge configurations have been generated by the Extended Twisted Mass Collaboration on the KNL (A2) Partition of Marconi at CINECA, through the Prace project Pra13\_3304 ``SIMPHYS''.
Inversions were performed using the DD-$\alpha$AMG solver~\cite{Frommer:2013fsa} with twisted mass support~\cite{Alexandrou:2016izb}.
\end{acknowledgments}

\appendix
\section{Details of Matching and RGR}
\label{app:matching_RGR}

The matrices entering the evolution and matching of Eq.~\eqref{eq:rgr} are
\begin{widetext}
\begin{align}
G(\xi)
&=
\begin{pmatrix}
1 & 0 & 0 & 0 \\
0 & 1 & 0 & 0 \\
-\dfrac{\xi^2}{5} & 0 & 1 & 0 \\
0 & -\dfrac{3\xi^2}{7} & 0 & 1
\end{pmatrix},
\,\,\,A(\mu^2,z^2)
=
\begin{pmatrix}
1 & 0 & 0 & 0 \\
0 & r^{-4/3} & 0 & 0 \\
0 & 0 & r^{-25/12} & 0 \\
0 & 0 & 0 & r^{-157/60}
\end{pmatrix},
\label{eq:GA_matrices}
\\[1em]
B(\xi,z^2)
&=
\begin{pmatrix}
0
& 0
& 0
& 0
\\[4pt]
0
& -a_s\left(-\dfrac{4}{3}l+\dfrac{14}{3}\right)
& 0
& 0
\\[4pt]
-\xi^2 a_s\left(\dfrac{5}{12}l-\dfrac{7}{6}\right)
& 0
& -a_s\left(-\dfrac{25}{12}l+\dfrac{47}{6}\right)
& 0
\\[4pt]
0
& -\xi^2 a_s\left(\dfrac{11}{20}l-\dfrac{53}{30}\right)
& 0
& -a_s\left(-\dfrac{157}{60}l+\dfrac{931}{90}\right)
\end{pmatrix},
\label{eq:B_matrix}
\end{align}
\end{widetext}
where $a_s$, $l$ and $r$ are defined in Eq.~\eqref{eq:rgr-shorthands}, with the coupling evaluated at the initial matching scale $\mu_0=2\kappa e^{-\gamma_E}/z$.

\section{Additional systematic tests}
\label{app:hyper}

\begin{figure}[t]
\centering
\includegraphics[width=\columnwidth]{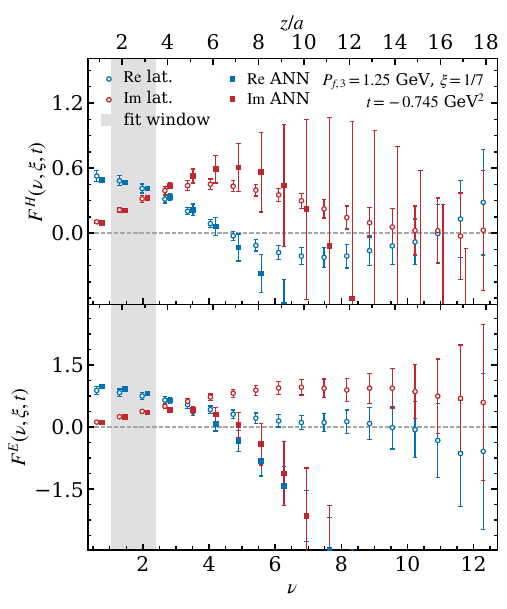}
\caption{Reduced ITDs at $P_{f,3}=1.25$~GeV, $\xi=1/7$ and $t=-0.745$~GeV$^2$ for all available Wilson line lengths $z/a=1,\dots,18$, compared with the predictions of the baseline fit. Open circles denote the lattice data and filled squares the ANN result, displaced horizontally by $\pm0.08$ in $\nu$. The top axis gives $z/a$, and the shaded band marks the separations $z/a=2,3$ that enter the training. The vertical range is set by the lattice data, so that ANN predictions at large $\nu$ lie outside the frame.}
\label{fig:app-itd-outside}
\end{figure}

In this appendix we collect two tests referred to in the main text. The baseline fit uses only $z/a=2$ and $3$, but its predictions can be compared with the lattice data at all available separations. Fig.~\ref{fig:app-itd-outside} shows this comparison for all separations $z/a=1,\dots,18$ at $P_{f,3}=1.25$~GeV, $\xi=1/7$ and $t=-0.745$~GeV$^2$, one of the kinematics of Fig.~\ref{fig:itd-fs-t075}, chosen because the lattice data are most precise at this momentum. The predictions for the neighboring separations $z/a=1$ and $z/a=4$, which are not part of the training, agree with the data within $1.3$ standard deviations in both channels, which shows that the network describes the $\nu$ dependence of the reduced ITDs around the fitted window rather than only the training points. In the $H$-type channel the agreement persists up to $z/a\simeq8$, whereas in the $E$-type channel the predictions deviate from the data by about two standard deviations from $z/a\simeq6$, that is $\nu\simeq4$, on, and the deviations grow rapidly at larger $\nu$ in both channels. This behavior is expected outside the short-distance region used in the fit: increasing \(z\) enhances the contribution of higher terms omitted by the truncation of Eq.~\eqref{eq:nu-expansion}, increases the power corrections neglected in the short-distance expansion, and increases the logarithms entering the perturbative Wilson coefficients, thereby reducing the reliability of perturbation theory. The comparison therefore supports restricting the extraction to the short-distance window \(z/a=2,3\). The test is complementary to the $\chi^2$ comparison of Sec.~\ref{sec:sys-z}, in which $z/a=1$ and $z/a=4$ are included in the training.

\begin{figure}[t]
\centering
\includegraphics[width=0.4943\columnwidth]{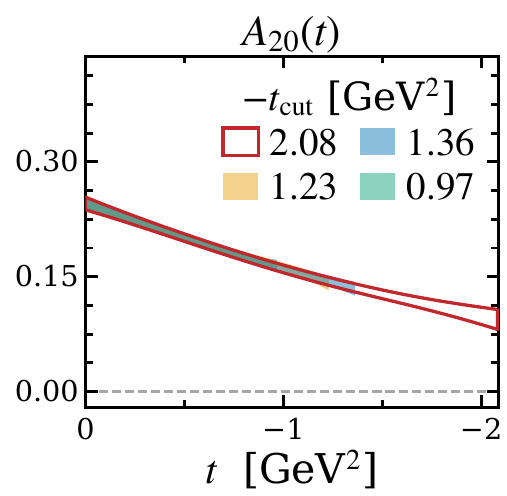}
\includegraphics[width=0.4943\columnwidth]{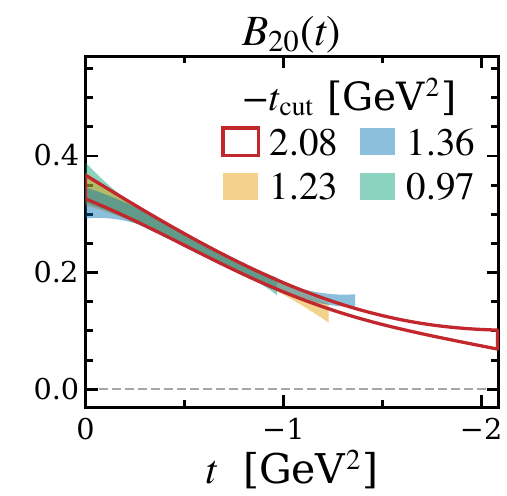}
\includegraphics[width=0.4943\columnwidth]{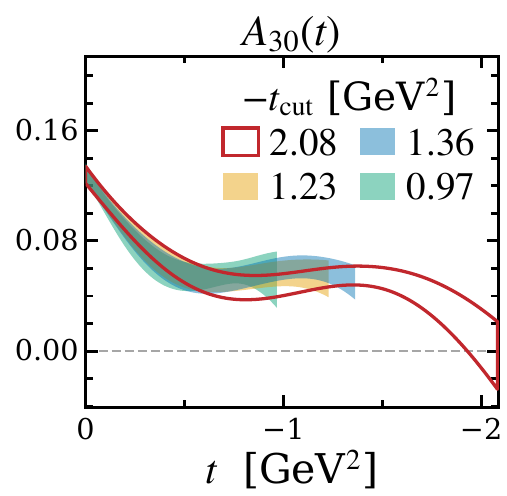}
\includegraphics[width=0.4943\columnwidth]{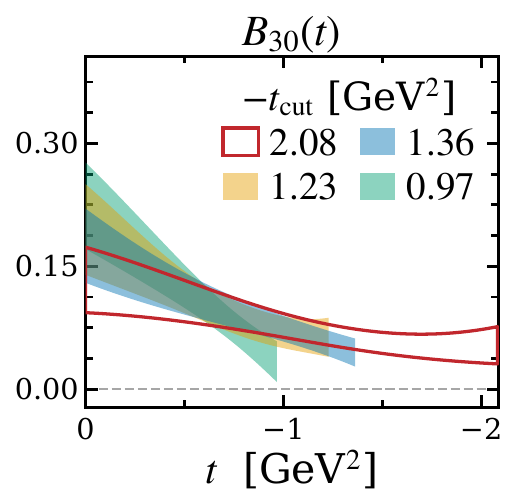}
\caption{Dependence of $A_{2,0}$, $B_{2,0}$ (upper panels) and $A_{3,0}$, $B_{3,0}$ (lower panels) on stronger cuts in the momentum transfer. The red outline is the baseline cut, which retains $90\%$ of the points ($-t_{\rm cut} = 2.08~\mathrm{GeV}^2$), and the filled bands correspond to cuts that retain $70\%$, $60\%$ and $50\%$ of the points ($-t_{\rm cut} = 1.36$, $1.23$ and $0.97~\mathrm{GeV}^2$, respectively). Each band extends only up to its own $|t_{\rm cut}|$, and the legend gives $-t_{\rm cut}$ in GeV$^2$. All other analysis choices are those of the baseline.}
\label{fig:app-tcut}
\end{figure}

Fig.~\ref{fig:app-tcut} extends the test of Sec.~\ref{sec:sys-tcut} to stronger cuts in the momentum transfer, retaining $70\%$, $60\%$ and $50\%$ of the kinematic points, i.e.\ removing the $30\%$, $40\%$ and $50\%$ of the points with the largest $|t|$. These cuts correspond to $-t_{\rm cut} = 1.36$, $1.23$ and $0.97~\mathrm{GeV}^2$, compared with $-t_{\rm cut} = 2.08~\mathrm{GeV}^2$ for the baseline, which retains $90\%$ of the points. As in Sec.~\ref{sec:sys-tcut}, the cuts are defined as fractions of the data set because of the nonuniform distribution of the $t$ values shown in Fig.~\ref{fig:kinematic-coverage}: the $20\%$ of the points removed in going from the baseline to the $70\%$ cut span $0.72~\mathrm{GeV}^2$ in $-t$, whereas the next $20\%$ span only $0.39~\mathrm{GeV}^2$. At $t=0$, the stronger cuts shift the four GFFs by at most $2.3$ standard deviations, the largest shift occurring for $B_{3,0}$ at the strongest cut.

\bibliographystyle{apsrev4-1}
\bibliography{reference}
\end{document}